\documentclass[aps,twocolumn,10pt,prc,floatfix,showpacs,preprintnumbers,amsmath,amssymb,nofootinbib,superscriptaddress]{revtex4-1}

\usepackage[normalem]{ulem}
\usepackage{wasysym}
\usepackage[dvipsnames]{xcolor}
\usepackage{graphicx}
\usepackage{overpic}
\usepackage{placeins}

\usepackage{tikz}
\usepackage{rotating}

\usepackage{dcolumn}    %
\usepackage{multirow, booktabs }
\usepackage{comment}
\usepackage{ulem}
\usepackage{siunitx}
\DeclareSIUnit{\nat}{nat}
\usepackage{setspace}
\usepackage{bm}
\DeclareSymbolFontAlphabet{\mathbb}{AMSb}
\DeclareMathAlphabet{\mathbbb}{U}{bbold}{m}{n}

\usepackage{lipsum}

\usepackage{float}

\usepackage{physics, mathtools}   %
\usepackage [ english ]{ babel }

\usepackage{bigstrut}
\usepackage{amsmath,amsfonts,amsthm,bm}
\usepackage{CJK}
\usepackage[pdfstartview=FitH,
            CJKbookmarks=true,
            bookmarksnumbered=true,
            bookmarksopen=true,
            colorlinks,
            linkcolor=blue,
            anchorcolor=blue,
            citecolor=blue,
            urlcolor=blue,
            ]{hyperref}
            
\usepackage{caption}
\usepackage{subcaption}
\renewcommand{\arraystretch}{1.2}

\usepackage{tikz}
\usetikzlibrary{calc}

\begin{document}	

\title{
    \Large \textbf{
    Null models for comparing information decomposition\\across complex systems}%
    }
\author{Alberto Liardi}
\email{a.liardi@imperial.ac.uk}
\affiliation{Department of Computing, Imperial College London, UK}
\affiliation{Cavendish Laboratory, Department of Physics, University of Cambridge, UK}%
\affiliation{\mbox{Center for Complexity Science, Department of Mathematics, Imperial College London, UK}}%

\author{Fernando E. Rosas}
\affiliation{\mbox{Center for Complexity Science, Department of Mathematics, Imperial College London, UK}}%
\affiliation{Sussex Centre for Consciousness Science and Sussex AI, University of Sussex, Brighton, UK}
\affiliation{Centre for Psychedelic Research, Department of Brain Sciences, Imperial College London, UK}

\affiliation{Centre for Eudaimonia and Human Flourishing, University of Oxford, UK}

\author{\mbox{Robin L. Carhart-Harris}}
\affiliation{Department of Neurology, University of California San Francisco, San Francisco, USA}

\author{George Blackburne}
\affiliation{Department of Experimental Psychology, University College London, UK}
\affiliation{Department of Computing, Imperial College London, UK}

\author{Daniel Bor}
\affiliation{Department of Psychology, University of Cambridge, UK}
\affiliation{Department of Psychology, Queen Mary University of London,
UK}

\author{Pedro A.M. Mediano}
\affiliation{Department of Computing, Imperial College London, UK}
\affiliation{Division of Psychology and Language Sciences, University College London, UK}

\begin{abstract}
\noindent
A key feature of information theory is its universality, as it can be applied to study a broad variety of complex systems. 
However, many information-theoretic measures can vary significantly even across systems with similar properties, making normalisation techniques essential for allowing meaningful comparisons across datasets. 
Inspired by the framework of Partial Information Decomposition (PID), here we introduce Null Models for Information Theory (NuMIT), a null model-based non-linear normalisation procedure which improves upon standard entropy-based normalisation approaches and overcomes their limitations. 
We provide practical implementations of the technique for systems with different statistics, and showcase the method on synthetic models and on human neuroimaging data. %
Our results demonstrate that NuMIT provides a robust and reliable tool to characterise complex systems of interest, allowing cross-dataset comparisons and providing a meaningful significance test for PID analyses.

\end{abstract}

\maketitle

\section*{Author summary}
How do complex systems process information? 
Perhaps more interestingly, when can we say two systems process information in the same way? 
Information-theoretic methods have been shown to be promising techniques that can probe the informational architecture of complex systems. 
Among these, information decomposition frameworks are models that split the information shared between various components into more elemental quantities, allowing for a more intuitive understanding of the system's properties. 
In the field of neuroscience, these measures are often used to gauge the differences between conscious states across health and disease. 
However, comparing these quantities across datasets is non-trivial, and simple normalisation techniques commonly employed have not been formally validated. 
In this work, we argue that such methods can introduce bias and result in erroneous conclusions, especially when the data under examination is significantly diverse.  
Our study sheds light on the origins of this issue, as well as its consequences and shortcomings. Moreover, it offers a rigorous procedure that can be employed to standardise these quantities, enabling more robust cross-dataset comparisons. 

\section{Introduction}  \label{sec:intro}

What are the emergent phenomena of a complex system, and how do we best quantify them? 
The interconnected and interdependent nature of the components of such systems, as well as their large numbers of degrees of freedom, %
are at the same time both their most compelling feature and their greatest challenge. 
These traits make complex systems unique in their ability to display emergent properties, in which the collective interactions can give rise to novel, unexpected, and self-organised phenomena that cannot be easily predicted by examining the individual parts in isolation. Examples of this abound in ecology~\cite{boccara2010modeling}, economics~\cite{arthur2009complexity}, neuroscience~\cite{girn2023complex}, and even in situations from everyday life such as traffic jams~\cite{kerner1996experimental}.

However, this richness in behaviour also poses great challenges to the investigation of their properties. 
A recent strategy to tackle this problem consists of describing the interactions between a system's elements by studying how information is routed and processed by each component of the system --- an approach known as \emph{information dynamics}~\cite{lizier2012local}. 
Within this field, a promising approach to unravel the relations among the constituents of complex systems is \emph{Partial Information Decomposition} (PID)~\cite{williams2010nonnegative}. This mathematical framework aims to characterise the interdependencies within a system by breaking down the information that two or more parts provide about another into unique, redundant, and synergistic contributions. 
A prominent feature of PID, inherited from information theory, consists in its broad applicability --- as the decomposition can be calculated on a wide variety of systems. This allows systematic comparisons of the interdependencies exhibited by different systems, and also between different states of the same system. 
In fact, PID has proven particularly useful to study artificial neural networks \cite{beer2015information, wibral2017partial, ince2017measuring, tax2017partial, proca2022synergistic}, gene regulatory systems \cite{chan2017gene, chen2018evaluating}, cellular automata \cite{finn2018quantifying, rosas2018information}, and neural dynamics \cite{luppi2020synergisticCore, varley2023partial, varley2023multivariate, luppi2024information, gatica2021high, gatica2022high}.

Despite its attractive features, PID %
suffers from some shortcomings that potentially limit its applicability. 
First, despite outlining the relations between the different modes of interdependency, the PID framework does not prescribe a unique functional form to calculate these quantities --- which has given rise to numerous proposals of how to best define these measures (see e.g.\ Refs.~\cite{williams2010nonnegative, bertschinger2013shared, griffith2014quantifying, harder2013bivariate, barrett2015exploration, ince2017measuring, james2018unique, griffith2014intersection, griffith2015quantifying, quax2017quantifying, rosas2020operational}). Additionally, 
results obtained through these various approaches can differ, potentially leading to seemingly contradictory conclusions (although, in many practical cases, several measures can yield qualitatively similar results~\cite{luppi2020synergisticCore, rosas2020operational}). 
Moreover, it is highly non-trivial to compare PID quantities across datasets, as their values are inherently dependent on the mutual information (MI) of the variables taken into consideration, a quantity that can vary greatly between systems with similar properties, and even more across various datasets. 
Hence, directly comparing PID atoms belonging to different distributions may yield results purely dictated by the difference in mutual information. To overcome this issue, a normalisation procedure is needed to quantify the amount of synergy, redundancy, and unique information, relative to the MI of the system.

Along this line, this paper introduces a novel methodological approach to allow more sensible PID comparisons of diverse systems' dynamics, alleviating both shortcomings mentioned above. 
Drawing from an established approach in network theory \cite{newman2018networks, vavsa2022null}, we introduce a null model technique for information-theoretic estimators that allows the comparison of quantities belonging to different information distributions, while also opening the way to more principled and effective cross-dataset comparisons. 
We also present a set of theoretical and empirical results that show that our method provides more robust conclusions than standard linear normalisations based on %
the MI of the system~\cite{kvalseth1987entropy, yao2003information, zamora2014line, vinh2009information}.  %
Furthermore, applications to real neural systems show that the proposed technique leads to consistent results across various datasets, even when using different %
PID formulations.

The rest of this article is structured as follows. We first describe the problem in Sec.~\ref{sec:key_intuitions}: focusing on a practical example, we show the non-linear behaviours of the PID atoms for different MI, discussing the limitations this poses for comparisons across systems. 
We then provide a solution in Sec.~\ref{sec:results} in the form of a null model for normalising PID results. 
After validating the method on synthetic models (Sec.~\ref{sec:synthetic}), we apply it to brain-scanning (magnetoencephalogram; MEG) data of subjects in altered states of consciousness (Sec.~\ref{sec:neural}). 
Finally, Sec.~\ref{sec:discussion} concludes with a discussion of implications and limitations. Methods are described in detail in Sec.~\ref{sec:methods}.

\section{The problem: comparing PID atoms between systems} \label{sec:key_intuitions}

This section presents the key problems tackled in this paper: the necessity of normalisation techniques for PID comparisons across different systems (Sec.~\ref{sec:keyint_norm}), and addressing the shortcomings of naive normalisation approaches that can lead to misleading results  (Sec.~\ref{sec:keyint_nmi}). 
We ground our intuitions on simple Gaussian systems, which let us investigate the behaviour of the PID atoms in a tractable manner.

\subsection{A simple example}
\label{sec:keyint_norm}

Given a system with two \textit{source} variables $X,Y$ and one \textit{target} variable $T$, PID proposes a decomposition of mutual information into four terms, or \textit{atoms}, as
\begin{multline} \label{eq:pid_tmi}
    I(X,Y; T) = \text{Red}(X,Y; T) + \text{Un}(X; T \setminus Y) \\
    + \text{Un}(Y; T \setminus X) + \text{Syn}(X, Y; T) ~ .
\end{multline}
The information associated to these atoms is commonly described as \emph{redundant} (information provided by both sources separately), \emph{unique} (provided by one source, but not the other), and \emph{synergistic} (provided only by both sources together, but neither of them in isolation). We refer to the quantity $I(X,Y; T)$ as the total mutual information, or TMI.

Unless otherwise specified, for the following analyses we employ the Minimal Mutual Information (MMI) PID~\cite{barrett2015exploration}, in which redundancy reads
\begin{equation}
    \text{Red}(X,Y;T) = \text{min}\Bigl(I(X;T), I(Y;T)\Bigr) \,,
\end{equation}
and the rest of the atoms follow from the defining PID equations (c.f.\ Eqs.~\eqref{eq:PID1} \eqref{eq:PID2} \eqref{eq:PID3}). 
However, our results also apply to other PID measures (App. \ref{app:neural_app}). 
We refer to Sec.~\ref{sec:methods} for a more in-depth discussion on this topic.

To develop our intuitions, consider two jointly Gaussian univariate sources $S = (X,Y)^T \sim \mathcal{N}(0,\Sigma_{S})$ and a one-dimensional target $T$ given by
\begin{equation} \label{eq:gauss_pid_distr}
T = A \,\begin{pmatrix} X \\Y \end{pmatrix} + \sqrt{g}\epsilon \,,
\end{equation}
where A is a fixed 1$\times$2 matrix of coefficients, ${\epsilon \sim \mathcal{N}(0,\Sigma_{\epsilon})}$ is a white-noise term, and $g\in \mathbb{R}^+$ is a parameter that determines the level of noise in the system. 
Intuitively, $A$ and $\Sigma_S$ describe \textit{how} the sources convey information about the target, while $g$ controls \textit{how much} information they provide. If $A$ and $\Sigma_S$ are fixed, the overall informational structure between $S$ and $T$ remains untouched --- although the TMI changes with different $g$. Therefore, we would expect that as $g$ increases the value of all atoms should decrease (as per the data processing inequality) -- but at least the overall qualitative character of the system (e.g.\ whether it is synergy- or redundancy-dominated) should not vary. As an illustration, in the rest of this section we adopt the following values:
\begin{equation} \label{eq:Gaussian_parameters}
    A = \begin{pmatrix} 0.5 & 0.5 \end{pmatrix} 
        \quad
    \Sigma_S = \begin{pmatrix}
                20 & 10\\
                10 & 20
                \end{pmatrix} 
        \quad
    \Sigma_{\epsilon} = 1 ~ .
\end{equation}

\begin{figure}[t]
\centering
    \includegraphics[width=\columnwidth]{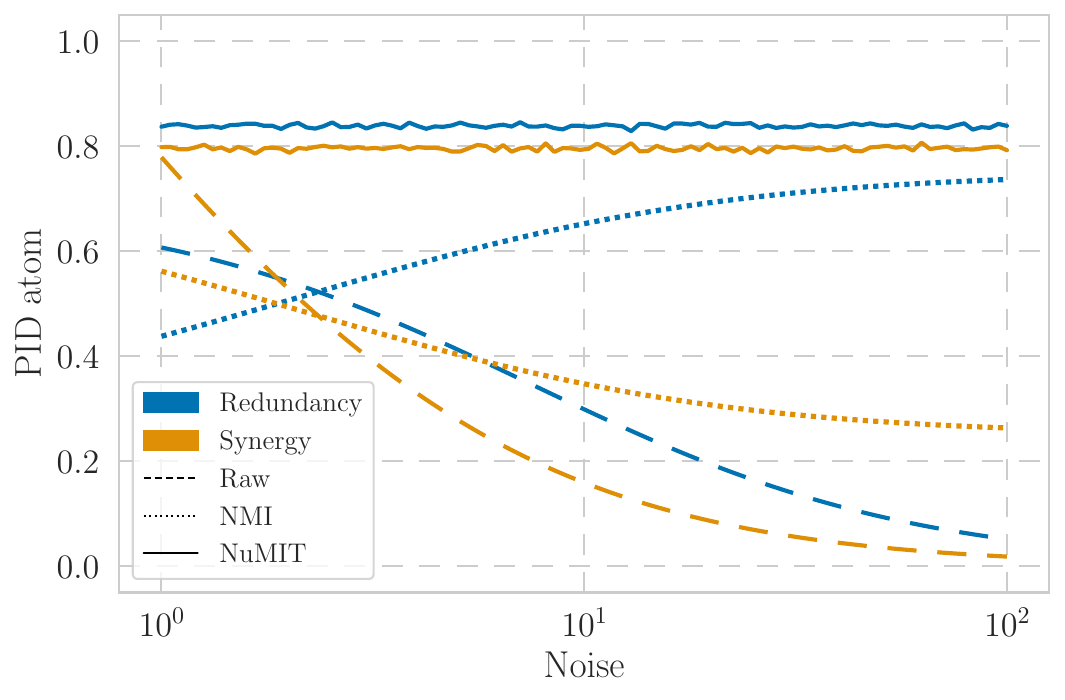}
\caption{Synergy and redundancy values for the bivariate Gaussian system given by Eqs. \eqref{eq:gauss_pid_distr} and \eqref{eq:Gaussian_parameters} for $g\in[1,100]$. Line styles represent raw atoms (dashed), atoms normalised by total mutual information (NMI, dotted), and atoms normalised by our proposed null-model procedure (NuMIT, solid).}
\label{fig:keyint_NMI}
\end{figure}

Calculating the MMI-PID for a range of values of $g$ shows that this is, surprisingly, not the case (see Fig.~\ref{fig:keyint_NMI}). Although raw values of PID atoms do decrease with $g$, the relative proportion between them changes radically, to the extent that (according to the raw PID values) higher $g$ causes the system to switch from being synergy- to redundancy-dominated. This counter-intuitive result is an example of a general and pervasive phenomenon, as similar behaviour is observed in higher-dimensional systems and using other PID measures (see Apps.~\ref{app:noise_sweep}, \ref{app:higher_dims}). 

We argue this behaviour is an important problem for the comparison of PID values across different systems, or across different conditions of the same system. For instance, if one were to observe data from two systems with the same source-target relationship ($A,\Sigma_S$) but different levels of signal-to-noise ratio ($g$), one may come to completely opposite conclusions. Therefore, this example illustrates the key problem we tackle in this paper: that raw values of PID atoms are often not comparable across systems with different values of TMI.

\subsection{Shortcomings of previous approaches} \label{sec:keyint_nmi}

The naive approach to obtain `normalised' PID atoms that do not depend so strongly on TMI is to simply divide each atom by TMI, i.e.
\begin{align}
    \text{Red}_\text{NMI}(X,Y; T) = \frac{\text{Red}(X,Y; T)}{I(X,Y; T)} ~ ,
\end{align}
and similarly for all other atoms. We refer to this procedure as \emph{Normalising by Mutual Information} (NMI).  NMI makes intuitive sense, is simple, and has the advantage that the resulting values can be understood as the proportion of the TMI that is contributed by each atom.

Unfortunately, however, results from the same Gaussian system as above show that NMI fails to solve the problem we aim to address (dotted lines in Fig.~\ref{fig:keyint_NMI}). NMI atoms still vary widely and show a switch from a synergy- to a redundancy-dominated decomposition as the noise increases. Hence, although this method seems a natural choice for PID normalisation, these results suggest that it might not be appropriate when comparing systems with large differences in TMI. 

To see why this happens, note that NMI entails a strong, yet implicit assumption: that the values of PID atoms grow linearly as TMI increases, and therefore that dividing by TMI would yield a value that does not depend on TMI itself. This would seem to make intuitive sense, since the atoms are indeed linearly related to TMI (c.f. Eq.~\eqref{eq:pid_tmi}). The key issue is that, as we show below, not all atoms grow in the same proportion, contradicting the implicit assumption of NMI.

\subsection{On the distribution of PID atoms} \label{sec:pid_distr_study}

If not linearly, how do the different atoms grow as the TMI increases? In this section we explore the relationship between the atoms and TMI, to show explicitly why NMI fails and to obtain important insights for our solution in Sec.~\ref{sec:results}.

We proceed by considering an ensemble of Gaussian systems with different $A, \Sigma_S$ that all yield the same TMI. Performing a PID decomposition on each provides a distribution of PID atoms that indicates the most likely values of synergy, redundancy, and unique information for systems with that specific value of TMI. 
In practice, we do this by randomly sampling each element of the coefficient matrix $A$ i.i.d. from a Gaussian distribution, $\Sigma_S$ from a Wishart distribution, and finding the value of $g$ that results in the desired TMI value. 
We refer to Sec.~\ref{sec:math_find} for the detailed description of the mathematical procedure followed.

\begin{figure*}[ht]
    \centering
    \hspace*{-10mm}\includegraphics[width=1.1\linewidth]{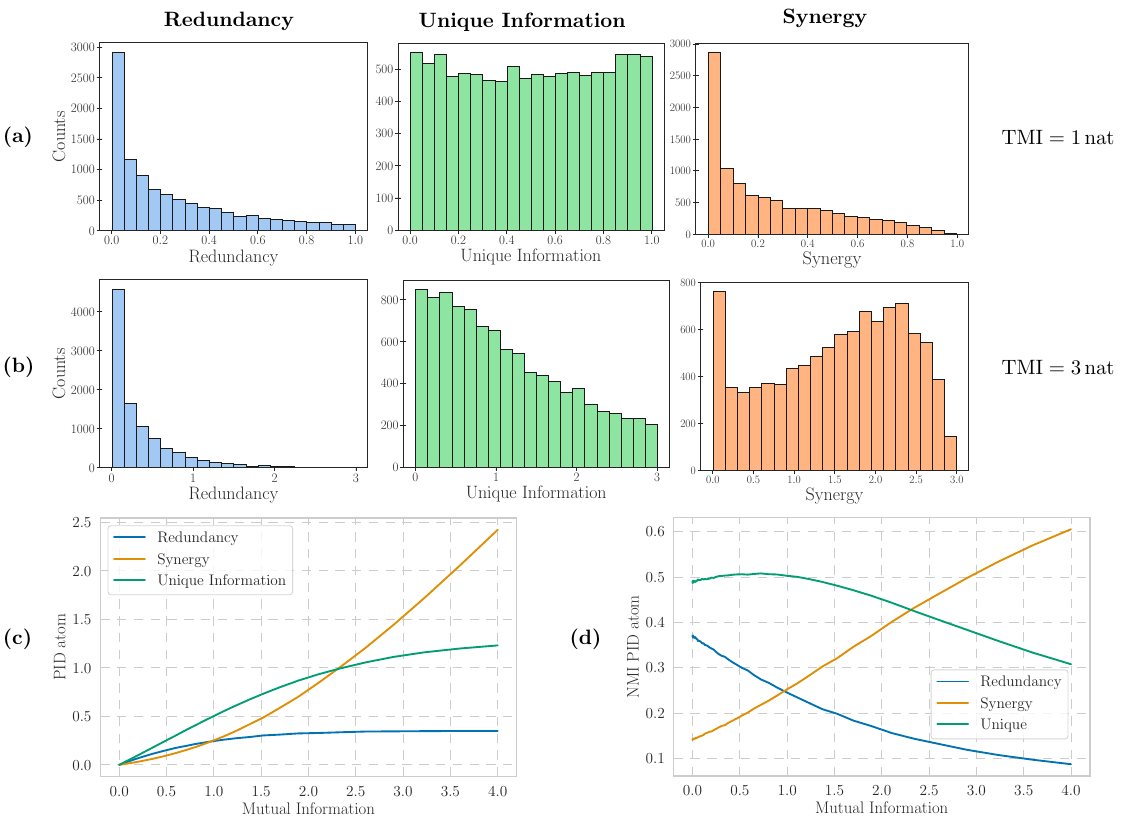}
\caption{(a) Distributions of redundancy, unique information, and synergy with MMI definition for TMI=1.0 nat and (b) TMI=3.0 nat. (c) PID-atom averages for different values of mutual information over random Gaussian systems with 2 univariate sources and a univariate target. (d) Same as (c) but with NMI-normalised PID atoms.
}
\label{fig:MI1_MI3_hist}
\begin{subfigure}{0.45\textwidth}
    \phantomcaption
    \label{fig:MI=1_pid}
\end{subfigure}
\hfill
\begin{subfigure}{0.45\textwidth}
    \phantomcaption
    \label{fig:MI=3_pid}
\end{subfigure}
\hfill
\begin{subfigure}{0.45\textwidth}
    \phantomcaption
    \label{fig:S2T1_all_PID}
\end{subfigure} 
\hfill
\begin{subfigure}{0.45\textwidth}
    \phantomcaption
    \label{fig:S2T1_all_PID_NMI}
\end{subfigure}
\end{figure*}%

To develop our intuitions, we begin by inspecting the distribution of each PID atom across Gaussian systems with two particular values of TMI (Figs.~\ref{fig:MI=1_pid} and \ref{fig:MI=3_pid}), drawing 10,000 samples from the ensemble in each case. 
From these, we observe that most Gaussian systems with $\text{TMI}=\SI{1}{\nat}$ are dominated by unique information, and high values of both synergy and redundancy are relatively rare. However, these distributions look very different among systems with $\text{TMI}=\SI{3}{\nat}$, which are clearly synergy-dominated, with large values of either redundancy or unique information becoming less likely. 
This indicates that the PID atoms behave qualitatively differently for different values of TMI, a feature that linear normalisation techniques are unable to capture.

As a further experiment, we repeat this procedure for a range of TMI values between 0 and \SI{4}{\nat}, calculating the mean of the distribution for each atom (Fig.~\ref{fig:S2T1_all_PID}). 
This shows that neither synergy, redundancy, nor unique information grows linearly with TMI. 
In fact, for systems with low TMI, on average, contributions to mutual information are mainly due to unique information and redundancy, whereas the synergistic component dominates for large TMI. 
The rapid growth of synergy for TMI above approximately \SI{3}{\nat} suggests that the information in systems with large TMI is mainly composed of synergistic contributions.\footnote{This can be easily understood in the limit of noiseless systems. If $T$ is a deterministic function of $X,Y$, then TMI is infinite, while the marginal mutual information of both $X$ and $Y$ remains finite -- thus, synergy must be infinite.} 
It is worth noting that increasing the number of sources (i.e.\ taking $\bm{X}, \bm{Y}$ multivariate) exacerbates this phenomenon, as the non-linear behaviour becomes more significant with the dimensionality of the sources (see Fig.~\ref{fig:MI_sweep_higher_dims} in App.~\ref{app:higher_dims}).

Taken together, the results in this section highlight some problems inherent to the comparison of PID atoms between systems with different TMI, and show why the naive NMI approach is not sufficient. Hence, a more sophisticated normalisation technique is needed.

\section{The solution: A null-model approach to normalise PID atoms}
\label{sec:results}

\subsection{Null model normalisation} \label{sec:math_find}

To address the problems raised in the previous section, we now present the central result of this work: a normalisation procedure that remains unaffected by the amount of noise in the system. 
In other words, given systems that only differ in the noise level but not in the statistical relationship between source and target variables, the desired method yields qualitatively similar PID decompositions. 

Inspired by the use of null models in network science and neuroscience~\cite{vavsa2022null,luppi2022dynamical}, the core idea of our method is to compare the PID atoms of the system of interest against those of \textit{an ensemble of all possible systems with the same TMI}.
In practice, we can operationalise this idea through the following algorithm: 

\begin{enumerate}
    \item Given the specific system under examination $p(S,T)$, calculate its TMI and perform the PID.
    \item Sample a null model $q_i(S,T)$ that has the same TMI as $p(S,T)$ but is otherwise random, and compute its PID.
    \item Repeat the previous step $N$ times for many sampled $q_i$, obtaining a null distribution of each PID atom.
    \item The relative amount of synergistic, unique, or redundant information of $p$ can be quantified by taking the quantile of the PID atoms of $p$ w.r.t.\ the null models $\{q_i\}_{i=1}^N$.
\end{enumerate}
\vspace*{-0.2cm}We call this approach \emph{Null Models for Information Theory} (NuMIT). 

Besides calculating the PID itself, the challenge of the algorithm above is step 2: taking a random sample from the set of probability distributions with a given TMI. 
We solve such a constraint by introducing a real parameter that, while not changing the underlying statistical structure of the model, can be tuned to yield the desired value of TMI.

\subsection{The Gaussian case} \label{sec:results_maths}
For simplicity, we show below the mathematical formulation of the method specifically for Gaussian systems. A formulation for autoregressive and discrete models is provided in Sec.~\ref{sec:VAR_null} and App.~\ref{app:discrete}, respectively.

Consider a multivariate Gaussian system $p(\bm{S}, \bm{T})$ consisting of two multivariate sources $\bm{X},\bm{Y}$ of dimension $d_X$ and $d_Y$, denoted jointly as ${\bm{S}=[\bm{X}^T \, \bm{Y}^T]^T}$ with dimension ${d_S = d_X + d_Y}$, such that ${\bm{S} \sim \mathcal{N}(0,\Sigma_{\bm{S}})}$, and be $\bm{T}$ a $d_T$-dimensional target variable given by
\begin{equation} \label{eq:gauss_pid_distr_mult}
    \bm{T} = A \bm{S}+ \sqrt{g} \bm{\epsilon} \,,
\end{equation}
where $A$ is a $d_T{\times}d_S$ coefficient matrix, $\bm{\epsilon}$ is a multivariate white-noise term $\bm{\epsilon} \sim \mathcal{N}(0,\Sigma_{\bm{\epsilon}})$, and $g$ is a non-negative real parameter that controls the noise strength.
Following from Eq.\ \eqref{eq:gauss_pid_distr_mult}, the covariance matrix of the target reads
\begin{equation} \label{eq:cov_T_gauss}
    \Sigma_{\bm{T}} = A \Sigma_{\bm{S}} A^T + g\Sigma_{\bm{\epsilon}} \, .
\end{equation}
Therefore, the total mutual information of $p$ can be written as
\begin{equation}
\begin{aligned}
    \text{TMI} \coloneqq I(\bm{S};\bm{T}) & = \frac{1}{2} \log{|\Sigma_{\bm{T}}|} - \frac{1}{2} \log{|g\Sigma_{\bm{\epsilon}}|} \\
    & = \frac{1}{2} \log{\left(\frac{|A \Sigma_{\bm{S}} A^T + g\Sigma_{\bm{\epsilon}}|}{|g\Sigma_{\bm{\epsilon}}|}\right)} \, . \label{eq:MI_gauss}
\end{aligned}
\end{equation}

We can create random systems $q_i(\bm{S}, \bm{T})$ by sampling both random coefficients $\tilde{A}$ and random covariance matrices $\tilde{\Sigma}_{\bm{S}}, \tilde{\Sigma}_{\bm{\epsilon}}$. One natural choice is to generate coefficient matrices with elements sampled i.i.d. from a Gaussian distribution $\tilde{A}^{ij} \sim \mathcal{N}(0,1)$, and covariance matrices from Wishart distributions $\tilde{\Sigma}_{\bm{S}} \sim W_{d_S}(\mathbbb{1}_{d_S{\times}d_S},d_S)$, ${\tilde{\Sigma}_{\bm{\epsilon}} \sim W_{d_T}(\mathbbb{1}_{d_T{\times}d_T},d_T)}$.

Finally, we can choose the parameter $g$ to obtain a null model $q_i$ that has a specific value of total mutual information.
We do this through an optimisation procedure, noting that from Eqs.\ \eqref{eq:cov_T_gauss} and \eqref{eq:MI_gauss} with straightforward calculations we obtain the function $f$ whose root determines the value of $g$:
\begin{equation}
    f(g) = \left|1+\frac{1}{g}\tilde{\Sigma}_{\mathbf{\bm{\epsilon}}}^{-1}\tilde{A}\tilde{\Sigma}_{\text{\textbf{S}}}\tilde{A}^T\right| - e^{-2 \cdot \text{TMI}} \, .
\end{equation}
It is straightforward to show that this is a monotonically decreasing function of the scalar parameter $g$, and therefore can be optimised with off-the-shelf tools.\footnote{We use the \texttt{fzero} solver in Matlab (R2021b, MathWorks, Natick, MA, USA), although we expect many other solvers to work too.} 
The sampled matrices $\tilde{A}, \tilde{\Sigma}_{\bm{S}}, \tilde{\Sigma}_{\bm{\epsilon}}$, together with the resulting value of $g$, determine a null system $q_i$. We can then calculate the PID atoms of $q_i$ and repeat this process multiple times to obtain the null distributions of each PID atom.

\subsection{Validation on synthetic systems} \label{sec:synthetic}

\begin{figure*}[ht]
    \centering
    \hspace*{-10mm}\includegraphics[width=1.1\linewidth]{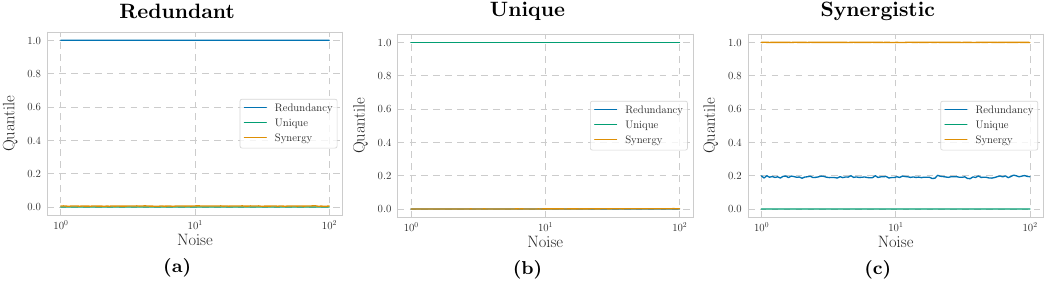}
\caption{Quantiles of the PID atoms for Gaussian models of Eq.\ \eqref{eq:gauss_pid_distr_mult}  for various noise levels $g\in[1,100]$. (a) Predominantly redundant system (Eq.\ \eqref{eq:max_red_sys}), (b) predominantly unique system (Eq.\ \eqref{eq:max_un_sys}), (c) predominantly synergistic system (Eq.\ \eqref{eq:max_syn_sys}).}
\label{fig:max_pid_sweep}
\begin{subfigure}{0.45\textwidth}
    \phantomcaption
    \label{fig:max_red}
\end{subfigure}
\hfill
\begin{subfigure}{0.45\textwidth}
    \phantomcaption
    \label{fig:max_un}
\end{subfigure}
\hfill
\begin{subfigure}{0.45\textwidth}
    \phantomcaption
    \label{fig:max_syn}
\end{subfigure}
\end{figure*}%

As a proof of concept, we apply the proposed method to three simple Gaussian systems with clear redundant, unique, or synergistic character, respectively, to check that our method yields the expected results. 
The aim is to both inspect the quantile of the predominant PID atom with respect to the null distribution, as well as test whether our model can capture the information structure of the system independently of the noise level.

As a first test, we expect redundancy to be highest when the two sources are highly correlated and are mixed via the same coefficients in $A$, thus providing the same information about $T$. For this purpose, we can choose:
\begin{equation} \label{eq:max_red_sys}
    A = \begin{pmatrix} 0.45 & 0.45 \end{pmatrix} 
        \,\,
    \Sigma_S = \begin{pmatrix}
                1 & 1-\delta \\
                1-\delta & 1
                \end{pmatrix} 
        \,\,
    \Sigma_{\epsilon} = 0.19 \,,
\end{equation}
where $\delta$ is a small positive number that ensures that $\Sigma_S$ is positive definite. The value of the covariance of $\epsilon$ was chosen so that for $g$=1 we have $\Sigma_T=1$. 
We set $\delta=10^{-4}$ and calculate PID atoms using the null-model procedure for a range of values of $g \in [1,100]$. Figure~\ref{fig:max_red} shows that results are precisely as expected: the normalised redundancy is approximately 1, while both synergy and unique information are close to zero.

Next, a predominantly unique system is one where a source is responsible for the majority of the total mutual information, while the other is uncorrelated with the target. We can achieve this by setting
\begin{equation} \label{eq:max_un_sys}
    A = \begin{pmatrix} 0 & 0.9 \end{pmatrix} 
        \quad
    \Sigma_S = \begin{pmatrix}
                1 & 0 \\
                0 & 1
                \end{pmatrix} 
        \quad
    \Sigma_{\epsilon} = 0.19 \,.
\end{equation}
We perform the same procedure as before and obtain Fig.~\ref{fig:max_un}, where the unique line is only the unique information brought by $Y$. Again, we observe that the normalised unique information is close to 1 and independent of the noise.

Finally, we focus on a predominantly synergistic system. This can obtained by highly anticorrelating the sources together while symmetrically correlating them with the target through $A$:
\begin{equation} \label{eq:max_syn_sys}
    A = \begin{pmatrix} 0.45 & 0.45 \end{pmatrix} 
        \,\,\,\,
    \Sigma_S = \begin{pmatrix}
                1 & -0.9 \\
                -0.9 & 1
                \end{pmatrix} 
        \,\,\,\,
    \Sigma_{\epsilon} = 0.998 \,,
\end{equation}
Fig.~\ref{fig:max_syn} reports the results of the NuMIT normalisation, confirming that the system is indeed maximally synergistic among the null models.

Hence, as well as producing the expected values, in all three cases above the normalised PID atoms indicate that the effects of the noise and mutual information on the decomposition are suppressed, satisfying the criteria outlined above.

\section{Case study: Information decomposition in cortical dynamics} \label{sec:neural}

The basic examples studied above show that the NuMIT-normalised PID atoms correctly quantify the information structure in simple systems. 
In this section, we analyse real-world brain activity data of subjects during altered states of consciousness to show that our method can reveal new insights about complex systems under study.

\subsection{Motivation and analysis set-up}

\begin{figure*}[ht]
    \centering
    \hspace*{-10mm}\includegraphics[width=1.1\linewidth]{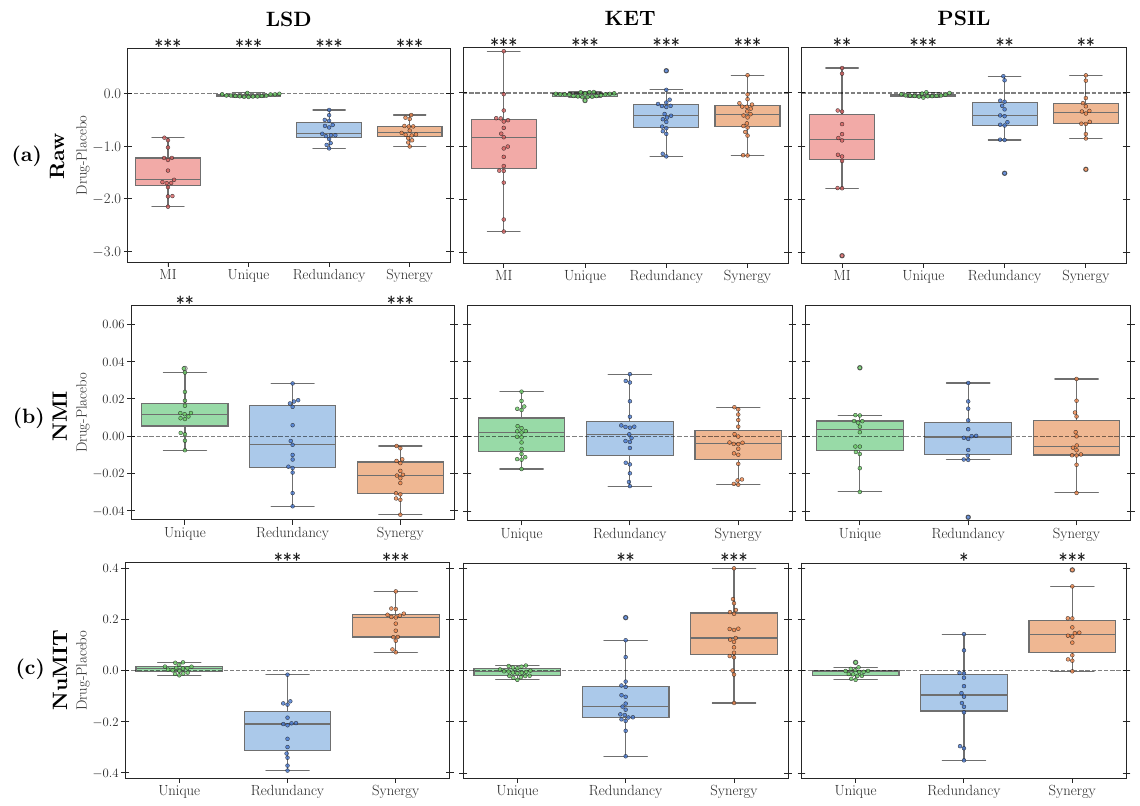}
\caption{PID-atom distributions for all subjects under different drugs and placebo effects using MMI definition. From left to right, results refer to LSD, ketamine, and psilocybin drugs. Panel rows represent (a) the raw values of PID atoms, (b) the NMI-normalised PID atoms, and (c) the NuMIT-normalised PID atoms. The dashed black lines are drawn at zero. (P-values calculated with a one-sample t-test against the zero-mean null hypothesis. *: $p<0.05$; **: $p<0.01$; ***: $p<0.001$).}
\label{fig:all_dp_MMI}
\begin{subfigure}{0.45\textwidth}
    \phantomcaption
    \label{fig:all_MMI_raw}
\end{subfigure}
\hfill
\begin{subfigure}{0.45\textwidth}
    \phantomcaption
    \label{fig:all_MMI_MI_norm}
\end{subfigure}
\hfill
\begin{subfigure}{0.45\textwidth}
    \phantomcaption
    \label{fig:all_MMI_Null_norm}
\end{subfigure}
\end{figure*}%

Information theory provides effective methods to assess important questions %
in the field of computational neuroscience, including the %
assessment of various aspects of cognition and consciousness. For instance, significant advances have been achieved in using complexity measures to characterise altered states induced by psychoactive substances~\cite{mediano2020effects, schartner2017relation, rajpal2022psychedelics}. 
Complementing these studies, here we explore the relationship between brain dynamics and conscious states by decomposing the information structure of such conditions with PID. 
One particularly interesting case is the change in neural activity elicited by psychedelic drugs like the serotonergic agonists LSD~\cite{carhart2016neural} and psilocybin~\cite{muthukumaraswamy2013broadband} and the NMDA antagonist ketamine~\cite{muthukumaraswamy2015evidence}. Previous works have reported a decrease in information flow due to these drugs~\cite{barnett2020decreased}, as well as a concomitant decrease in TMI between past and future~\cite{mediano2023spectrally} -- prompting the question of whether, or to what extent, the decrease in information flow can be explained by the decrease in TMI.

Addressing a similar question but in the context of PID, we analyse resting-state magnetoencephalography (MEG) recordings of subjects under the effects of different psychedelic drugs --- ketamine (KET) (N=19) \cite{muthukumaraswamy2015evidence}, LSD (N=15) \cite{carhart2016neural}, and psilocybin (PSIL) (N=14) \cite{muthukumaraswamy2013broadband} --- and matching placebo recordings. 
More details about the datasets and pre-processing pipeline are provided in Sec.~\ref{sec:neural_data}, the open data repository~\cite{megdata}, and in the original studies~\cite{muthukumaraswamy2015evidence, carhart2016neural, muthukumaraswamy2013broadband}. 
Since the data is in the form of multivariate time series of brain activity, we model them using a Vector Autoregression (VAR) process. VAR models are powerful and versatile tools in the time series modelling literature, and they are particularly attractive for information-theoretic analyses due to their tractability~\cite{faes2015information, faes2016information, faes2017multiscale}. NuMIT normalisation presented in Sec.~\ref{sec:math_find} can be naturally extended to VAR models -- for further mathematical details of the framework and normalisation procedure see Secs.~\ref{sec:VAR} and~\ref{sec:VAR_null}.

\subsection{Null-model normalisation reveals higher synergy under psychedelics}

\begin{figure*}[ht]
    \centering
    \hspace*{-10mm}\includegraphics[width=1.1\linewidth]{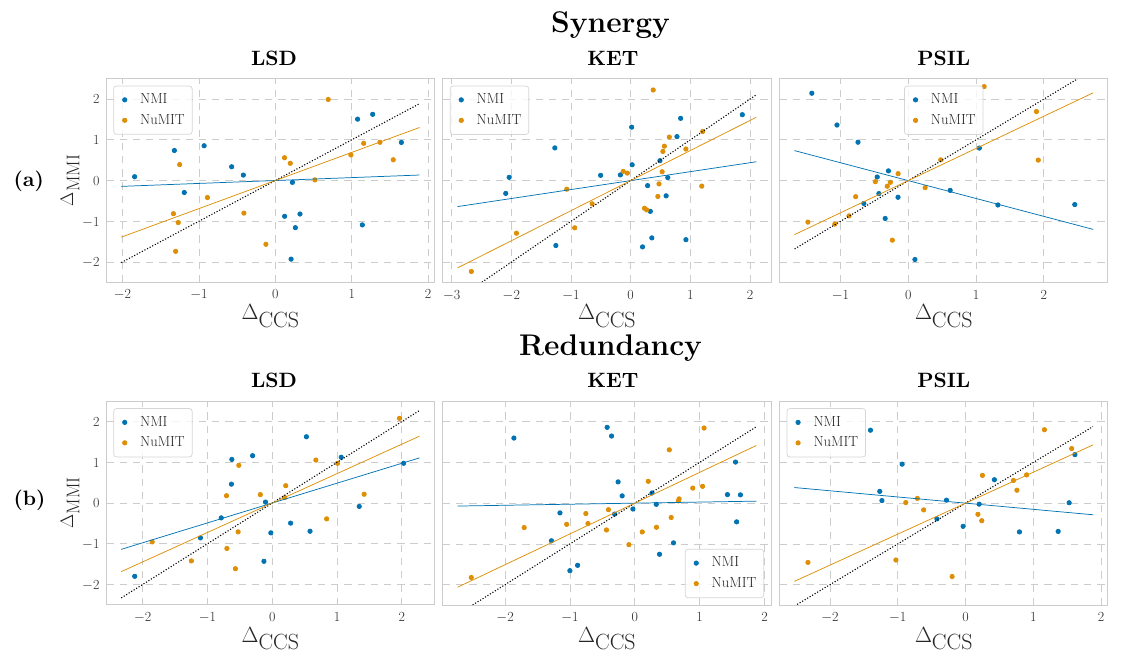}
\caption{Regression models for NMI- and NuMIT-normalised (a) synergies and (b) redundancies between $\text{CCS}$ and MMI PID definitions, for LSD, ketamine, and psilocybin drugs. $\Delta$ indicates the differences between drug and placebo in PID atoms obtained with either PID (MMI or CCS).}
\label{fig:mult_regr}
\begin{subfigure}{0.45\textwidth}
    \phantomcaption
    \label{fig:mult_regr_lsd_iccs_mmi}
\end{subfigure}
\hfill
\begin{subfigure}{0.45\textwidth}
    \phantomcaption
    \label{fig:mult_regr_lsd_iccs_mmi_red}
\end{subfigure}
\end{figure*}%

Our analysis starts from the time series of 90 brain regions source-reconstructed from 271 MEG channels according to the AAL atlas~\cite{tzourio2002automated}.
For every subject, drug (ketamine, LSD, or psilocybin), and condition (drug or control), we sample 1000 sets of 10 random regions, %
split them into two subsets, and fit a VAR(1) model to calculate their PID using their past state as sources and their joint future state as target. Then, we calculate the raw PID atoms, as well as those normalised using NMI and NuMIT. 
Finally, we average each PID atom across all regions to produce a single value for each subject, drug, condition, and normalisation procedure. More details of this procedure can be found in Sec.~\ref{sec:PID_var_method}.

The first notable result is that all raw PID atoms decrease in the psychedelic state compared to placebo, consistently across all drugs (Fig.~\ref{fig:all_MMI_raw}). This is expected given the strong (and previously reported\footnote{Although we do not know of any reports of TMI \textit{per se}, there are many reports of increased entropy rate (often estimated through Lempel-Ziv complexity) under psychedelics~\cite{mediano2023spectrally,schartner2017relation}. Since negative entropy rate and TMI need to sum up to the entropy of the channel, which is normalised, an increase in one is equivalent to a decrease in the other.}~\cite{mediano2023spectrally}) decrease of TMI --- since the sum of all atoms decreases between conditions, it makes sense that the atoms that constitute TMI decrease too. Atoms normalised by NMI (Fig.~\ref{fig:all_MMI_MI_norm}) are largely non-significant, with the few significant changes not consistent across drugs. These results call for a suitable normalisation of PID atoms to enable a more meaningful comparison.

Accordingly, we applied our null model normalisation to these findings (Fig.~\ref{fig:all_MMI_Null_norm}). With this technique, we now observe a very consistent pattern of increased synergy and decreased redundancy in the psychedelic state across all three drugs. Therefore, although there is overall less synergy in the psychedelic state, there is in fact more synergy than one would expect given the brain's TMI. Although these findings still need further interpretation in the broader context of psychedelic neuroimaging, they show the benefits of our proposed normalisation across diverse datasets.

\subsection{Null-model normalisation increases consistency across PID measures} \label{sec:res_mult_regr}

One common concern in experimental applications of PID is the lack of consensus on a one-size-fits-all PID measure. Accordingly, there has been a proliferation of PID measures in the literature (see e.g.\ Refs.~\cite{barrett2015exploration,ince2017measuring,harder2013bivariate,griffith2014quantifying,james2018unique} as a non-exhaustive list), raising concerns that different measures may yield conflicting results on the same dataset. Here we argue that our null model normalisation can alleviate this concern, by showing that a proper normalisation makes the results of different PID measures more consistent. We focus here on the comparison between MMI and two other PID methods: Common Change in Surprisal (CCS)~\cite{ince2017measuring}, and Dependency Constraints (DEP)~\cite{james2018unique} (App.~\ref{app:mult_regr_app}).

As a first analysis, we follow the same procedure as before and compute raw, NMI, and NuMIT-normalised CCS results (Fig.~\ref{fig:all_dp_Iccs} in App.~\ref{app:neural_app}). 
A comparison between CCS and MMI shows that although the distributions of the raw and NMI PID atoms are quite different between both measures, the null-normalised atoms give substantially more consistent patterns. 
To make this observation more rigorous, we investigate whether CCS and MMI yield consistent conclusions about the effect of psychedelics on brain activity -- i.e.\ whether the difference between drug and placebo conditions on each subject is the same across both PID measures.

For this purpose, we construct a regression model that assesses the correlations of both NMI and the NuMIT-normalised atoms across MMI and CCS PIDs, while also providing a p-value to test whether the differences between the two distributions are statistically significant. 
A complete description of the model and the parameters used is reported in Sec.~\ref{sec:meth_mult_regr}.

Results in Tab.~\ref{tab:mult_regr_Iccs_MMI} and Fig.~\ref{fig:mult_regr} show that the NuMIT normalisation increases the correlation between the results of CCS and MMI definitions. Although this effect was only considered statistically significant by our test in some of the cases, the null model normalisation still increased the correlation between measures to nearly $0.7$ and above in all cases. Results with DEP indicate similar trends (App.~\ref{app:mult_regr_app}), with slightly smaller correlations with MMI and much higher with CCS (Tabs.~\ref{tab:mult_regr_Idep_MMI} and~\ref{tab:mult_regr_Iccs_Idep}). 
Thus, after suitable normalisation, all three PID measures qualitatively agree on the effects of psychedelics on brain activity, boosting our confidence in the results of the analyses.

\begin{table}[]
\begin{tabular}%
    {
    >{\centering}p{0.2\columnwidth} 
    >{\centering}p{0.2\columnwidth}
    >{\centering}p{0.15\columnwidth}
    >{\centering}p{0.15\columnwidth}
    >{\centering}p{0\columnwidth} 
    >{\centering\arraybackslash}p{0.15\columnwidth}
    }
\multirow{2}{*}{PID atom} & \multirow{2}{*}{Drug} & \multicolumn{2}{c}{Correlations} &  & \multirow{2}{*}{p-value} \\ \cline{3-4}
&  & NMI & NuMIT &  &  \\ \hline
\multirow{3}{*}{Synergy} 
& LSD & 0.071 & 0.690 &  & 0.08 \\
& KET & 0.220 & 0.738 &  & 0.08 \\
& PSIL & -0.439 & 0.790 &  & 0.001 \\ \hline
\multirow{3}{*}{Redundancy} 
& LSD & 0.488 & 0.723 &  & 0.45 \\
& KET & 0.026 & 0.755 &  & 0.02 \\ 
& PSIL & -0.152 & 0.760 &  & 0.01 \\ \hline
\end{tabular}
\caption{Pearson correlations and p-values of the correlation increase from NMI to NuMIT, for synergy and redundancy and for all drugs (MMI-CCS comparison).}
\label{tab:mult_regr_Iccs_MMI}
\end{table}

\section{Discussion}  \label{sec:discussion}

We proposed a new methodological framework to quantify the significance of structural measures of information in a complex system. %
Our approach is based on null models, a method that is widely employed in the analyses of complex networks, but still has not been widely adopted in the context of information-theoretic analyses. 
This technique has proven useful in understanding non-trivial structures and dynamics, often indicative of meaningful organisation within complex systems.

In this paper we applied this philosophy to the Partial Information Decomposition framework. 
The rationale behind this approach is to employ null models to eliminate the intrinsic dependency of information-theoretic quantities on the mutual information of the system, opening the way to comparisons of PID atoms across datasets that show great informational variability.

\subsection{Summary of findings}
We first focused on a simple Gaussian model and analysed the effect of the common technique of normalising by mutual information (NMI) each PID atom. 
Results showed that the distributions of PID components present heavy non-linear dependencies on the total mutual information of the system, contradicting the hidden linearity assumptions behind NMI. %
As a solution, we suggested a new normalisation method based on the construction of null models. 
This was performed by generating an ensemble of all systems with the same total mutual information as the system under study, and taking the distribution of PID atoms in this ensemble as null distribution. 
Finally, we took the respective quantiles of each atom on its null distribution as universal estimators of how redundant, synergistic, and unique the observed system is.

After outlining the mathematical foundation of the technique, we proved its advantages empirically by applying it to synthetic data and real systems. 
Direct analyses of neural data showed that NuMIT provided consistent and significant findings, specifically revealing higher synergistic contributions in drug-induced conditions and supporting previous studies in the literature~\cite{luppi2024synergistic}. 
Such results could not have been obtained with linear normalisation techniques, underlining the importance and necessity of taking into account non-linear PID atoms' behaviours.

Moreover, a key aspect of an effective normalisation procedure is that it is robust to different ways of calculating PID. In the case of our model, the observations above were validated with multiple PID definitions, showing that the null model approach allows for a more coherent characterisation of the information structure of the system. 
In fact, a direct comparison between atoms computed with different PIDs showed that the correlations between null-normalised atoms across different PID methods are higher than those obtained with the NMI technique, being consistent for the majority of the methods and drugs.

\subsection{Neural null models}

The concept of null models has been introduced in network theory \cite{maslov2002specificity} as a tool to assess the significance of observed patterns in complex networks. These techniques are designed to generate randomised or controlled versions of a given graph while preserving certain structural characteristics, such as the number of nodes and edges, but introducing randomness in other aspects, such as the arrangement of connections \cite{newman2004finding}. 
By comparing network metrics derived from real-world networks to those from null models, it is possible to determine whether the observed features are the result of genuine underlying properties or merely a consequence of random chance. 
Therefore, null models provide a baseline for statistical inference in network analysis, helping identify patterns and properties such as node degree distributions~\cite{mahadevan2006systematic}, community structure~\cite{cazabet2017enhancing} and detection~\cite{sarzynska2016null, paul2021null}, assortativity~\cite{pastor2001dynamical, foster2010edge}, and more.

In this work we provided a procedure to construct null models for information theory, allowing more rigorous and mathematically robust comparisons and characterisations of information-theoretic quantities. 
While this technique offers a foundation for better normalisations, it is important to note that the effectiveness of null model comparisons heavily depends on the suitability of the generated null systems to the specific problem under study. In other words, these null models must resemble the original data in some meaningful way, as they could otherwise differ significantly from any real system and be of poor practical use. Therefore, the choice of null models needs to be optimised for each complex system and each experimental hypothesis under consideration. 

In our scenario, not knowing a biologically meaningful null model could pose a possible limitation to the study, as real structures of neural null models remain unknown due to the exceptional complexity of the brain. 
For instance, considering null models significantly diverse from the real system could prevent a proper characterisation of the feature of interest, which might have been possible if only realistic null systems were taken into account.  
Nevertheless, if clear and significant results arise, such as the ones presented in Sec.~\ref{sec:results}, these are robust indicators of structural information diversities. %

\subsection{Final remarks}

The results reported in this paper suggest that null-model techniques have great potential for enabling meaningful comparisons of information-theoretic analyses between systems. 
This is relevant not only for neural systems but also for a wide range of fields, as variations in mutual information can be observed from financial and stock markets \cite{fiedor2014networks, guo2018development}, to ecological systems with different sizes and evolution dynamics \cite{nicoletti2022mutual, nicoletti2024information,rajpal2023quantifying}. 
Nonetheless, further studies should examine the structure of meaningful null models in these domains and their applications to various real-world datasets. 

On the technical side, possible generalisations may entail the development of the technique for other statistical models, e.g.\ moving-average or state-space models, and more refined TMI-based information frameworks, like Integrated Information Decomposition ($\Phi$ID) \cite{mediano2019beyond}. 
Additionally, the generality of the procedure here performed on TMI enables applications to other core information-theoretic quantities and their decompositions, such as the joint entropy in Partial Entropy Decomposition (PED) \cite{ince2017partial, finn2020generalised, varley2023partial} and the KL-divergence in Generalised Information Decomposition (GID) \cite{varley2024generalized}. 

Future investigations could examine non-Markovian dynamics using VAR(p) models with $p>1$, which, although already theoretically possible, in practice involves a higher computational load, but might shed light on long-range causal effects within brain activity. 
However, an important limitation of these models is that they only describe systems with linear dynamics, even though nonlinear relationships are essential to understanding and emulating the behaviour of complex systems \cite{kaplan2012understanding, fuchs2014nonlinear}. 
Thus, further studies should be devoted to exploring more general dynamical processes \cite{jones1978nonlinear}. 

Overall, we hope the presented findings may foster further investigations on the potential of null models for complementing information-theoretic methods. 
To encourage the usage of these methods in the information theory community, we provide the code for the null model normalisation in a publicly available GitHub repository \href{https://github.com/alberto-liardi/NuMIT}{here}.

\section{Materials and Methods}  \label{sec:methods}

\subsection{Partial Information Decomposition}  \label{sec:PID} 
Information theory is largely based upon the notion of Shannon entropy, which quantifies the average information content in a random variable~\cite{mackay2003information}. In other words, the entropy of the stochastic variable $X$ accounts for the information we gain (on average) about the system after $X$ is measured, and it's defined as
\begin{equation} \label{eq:def_entropy}
    H(X) = -\sum_x p(x) \log{p(x)} \, ,
\end{equation}
where $x$ indicates the possible outcomes of $X$. 
However, entropy alone does not capture the relations between variables in a system. In contrast, mutual information (MI) is a measure of the average information shared between two variables $X,Y$, and can be interpreted as the reduction in uncertainty about $X$ given the knowledge of $Y$. Its definition reads 
\begin{equation} \label{eq:MI_entr_def}
    I(X;Y) = H(X) - H(X|Y) \, ,
\end{equation}
where $H(X|Y)$ is the entropy of $X$ conditioned over $Y$
\begin{equation} \label{eq:cond_entr_def}
\begin{aligned}
    H(X|Y) & =  -\sum_y p(y)\,H(X|y) \\
    & = -\sum_{x,y} p(x,y) \log{p(x|y)}\, .
\end{aligned}
\end{equation}
However, in a complex system constituted of many elements, mutual information cannot capture high-order interactions as it only describes pairwise relations. 
Considering a 3-variable system, with sources $X,Y$ and a target $T$, the framework of PID solves this limitation by decomposing pairwise MI into three kinds of quantities: unique information (Un), redundancy (Red), and synergy (Syn):
\begin{align}
    I(X;T) =~ &\text{Un}(X; T \setminus Y) + \text{Red}(X,Y; T) \label{eq:PID1} \\
    I(Y;T) =~ &\text{Un}(Y; T \setminus X) + \text{Red}(X,Y; T) \label{eq:PID2} \\
    \begin{split}
        I(X,Y; T) =~ &\text{Red}(X,Y; T) + \text{Un}(X; T \setminus Y) \\
    ~ &+ \text{Un}(Y; T \setminus X) + \text{Syn}(X, Y; T) ~ \label{eq:PID3},
    \end{split}
\end{align}
where $I(X,Y;T)$ is the joint mutual information that $X,Y$ provide about $T$, whereas $I(X;T)$, $I(Y;T)$ are the marginal mutual information of $X$ and $Y$ respectively.

However, the PID equations do not provide a unique solution to calculate such quantities, as they form an underdetermined system of three equations and four unknowns (Eqs. \eqref{eq:PID1}, \eqref{eq:PID2}, \eqref{eq:PID3}), and various studies have been devolved in finding a suitable expression~\cite{williams2010nonnegative, bertschinger2013shared, griffith2014quantifying, harder2013bivariate, barrett2015exploration, ince2017measuring, james2018unique, griffith2014intersection, griffith2015quantifying, quax2017quantifying, rosas2020operational}. 
In this work we mainly focused on the \textit{Minimum Mutual Information} (MMI) definition proposed by Barrett~\cite{barrett2015exploration}, who proved that, for 3-variable Gaussian systems, seemingly different versions of redundancy previously suggested in the literature reduce to the simpler and more intuitive expression:
\begin{equation}
\begin{aligned}
    \text{Red}(X,Y;T) & = \text{Red}_{\text{MMI}}(X,Y;T) \\ 
    & = \text{min}\Bigl(I(X;T), I(Y;T)\Bigr). \label{eq:Barrettred}
\end{aligned}
\end{equation}
Although this is the central definition used in the work, for the analyses on real data we also considered the \textit{Unique Information via Dependency Constraints} (DEP) by James \cite{james2018unique} and the \textit{Common Change in Surprisal} (CCS) definition by Ince \cite{ince2017measuring}, showing how our method yields consistent results with all three definitions. 

Although in theory the PID decomposition can be applied to a system of $N$ sources, in practice this leads to a super-exponential growth of PID atoms and intractable computational loads for high values of $N$. 
Therefore, throughout this work, we considered an arbitrary number of sources and partitioned them into two multivariate variables $\bm{X}$ and $\bm{Y}$, then employed the 3-variable PID described above. 

Hence, studying a system's information dynamics with PID entails computing the entropies of the variables in the system, then the mutual information between sources and target, and finally proceeding with the PID decomposition.

\subsection{Vector autoregression model}  \label{sec:VAR}
Vector AutoRegression (VAR) is a statistical model employed to study multivariate time series. These models allow for tractable analyses of the interactions between variables in a complex system~\cite{barrett2010multivariate}, and has found broad applications in economics~\cite{sims1980macroeconomics}, statistics~\cite{hatemi2004multivariate}, and social sciences~\cite{box2014time}. 
VAR models have already been applied to the study of biological systems \cite{faes2015information, faes2016information}, offering more robust and easier estimations of the covariance of the system compared to direct calculations from the raw signal. 
In this work, we employed this framework to interpret multivariate time series coming from MEG data, employing VAR to capture the statistical relations between different regions of the brain (Sec.~\ref{sec:results}).

The framework consists of a set of linear equations in which the evolution of the system follows a deterministic growth of the past states, plus a noise term. 
Considering an $n$-dimensional time series $\mathbf{X}_t$, the VAR equation can be written as:
\begin{equation}
    \mathbf{X}_t = \sum_{l=1}^p \text{A}_l \mathbf{X}_{t-l} + \boldsymbol{\eta}_t \, , 
    \label{eq:VAR_eq3}
\end{equation}
where we introduced the shorthand matrix notation
\begin{equation*}
    \mathbf{X}_i = \begin{pmatrix} X^1_i \\ \vdots \\X^n_i \end{pmatrix}, \quad \quad \quad
    \boldsymbol{\eta}_i = \begin{pmatrix} \eta^1_i \\ \vdots \\ \ \eta^n_i \end{pmatrix} , 
\end{equation*} 
where $p$ is a positive integer, $\boldsymbol{\eta}$ a multivariate white noise term, and $A_l$  the $n{\times}n$ evolution matrix that contains the $l$-th lag VAR coefficients.
In neuroscience, these are also called \textit{effective connectivity matrices}, as they describe the effective interdependencies between different regions of the brain. Notice how the model only takes into account the last $p$ steps of the time series, ignoring further past states. For this reason, $p$ is named the \textit{model order} and the system is called VAR(p).

Conducting an information dynamics analysis involves studying the covariances between the time-lagged variables. These appear as elements of the autocovariance matrices $\Gamma_k \equiv \mathbb{E}[\mathbf{X}_t\, \mathbf{X}_{t-k}^\mathrm{T}]$, which can be computed through the Yule-Walker equations \cite{yule1927vii, walker1931periodicity}: 
\begin{equation}
   \Gamma_k = \sum_{l=1}^p \text{A}_l \Gamma_{k-l} + \delta_{k0} V, \quad k=0, ..., p \,, 
   \label{eq:YW}
\end{equation}
where $\delta_{k0}$ is the Kronecker delta, and V the residual covariance, i.e.\ the covariance matrix of the white noise distribution. 
Thus, Eq.~\eqref{eq:YW} provides a recipe to obtain $\Gamma_k$ matrices through a recursive relation.
However, instead of solving Eq.~\eqref{eq:YW} directly, we can introduce the following quantities
\begin{gather}
    \mathbf{S}_t =
    \begin{pmatrix}
    \bm{X}_t \\ \bm{X}_{t-1} \\ \vdots \\ \bm{X}_{t-p+1}
    \end{pmatrix} ,\,\, %
    \bm{\epsilon}_t = 
    \begin{pmatrix}
    \bm{\eta}_t \\ 0 \\ \vdots \\ 0
    \end{pmatrix}, \\ \vspace{0.3pt} %
    \mathbf{A} = 
    \begin{pmatrix}
    \text{A}_1 & \text{A}_2 & \cdots & \text{A}_{p-1} & \text{A}_p \\
    \mathbbb{1} & 0 & \cdots & 0 & 0 \\
    0 & \mathbbb{1} & \cdots & 0 & 0 \\
    \vdots & \vdots & \ddots & \vdots & \vdots \\
    0 & 0 & \cdots & \mathbbb{1} & 0 
    \end{pmatrix} 
    \label{eq:A^p} , \\ \vspace{0.3pt}
    \mathbf{\Gamma} =
    \begin{pmatrix}
    \Gamma_0 & \Gamma_1 & \cdots & \Gamma_{p-1} \\
    \Gamma_1^\mathrm{T} & \Gamma_0 & \cdots & \Gamma_{p-2} \\
    \vdots & \vdots & \ddots & \vdots \\
    \Gamma_{p-1}^\mathrm{T} & \Gamma_{p-2}^\mathrm{T} & \cdots & \Gamma_0 
    \end{pmatrix} ,
    \label{eq:gamma^p}
\end{gather}
and formally rewrite Eq.~\eqref{eq:VAR_eq3} as a VAR(1) model:
\begin{equation}
    \mathbf{S}_t = \mathbf{A} \mathbf{S}_{t-1} + \bm{\epsilon}_t .
\end{equation}
From here it follows that $\mathbf{\Gamma}$ needs to satisfy the discrete Lyapunov equation \cite{parks1992lyapunov} given by
\begin{equation}
    \mathbf{\Gamma} = \mathbf{A} \mathbf{\Gamma} \mathbf{A}^\mathrm{T} + \text{\textbf{W}} ,
    \label{eq:Lyapnuov}
\end{equation}
where \textbf{W} is a block matrix with residual covariance V in the first entry and zeros elsewhere.   

Hence, once $A_k$ and $V$ are known, it is possible to obtain the autocovariance matrices from Eq.~\eqref{eq:Lyapnuov} and then compute the full covariance matrix of the system.

\subsection{Estimating PID from VAR models} \label{sec:PID_var_method}
In this work we studied brain activity employing the PID framework and VAR models. 
Here we provide the technical details of the procedure.

For each subject in each specific condition, the analysis proceeds from sets of 10 brain regions randomly chosen from the dataset, considering their time series across 50 random epochs. We further randomly split these 10 brain regions into two sets, and consider their past states as sources $X,Y$ and their joint future state as target $T$.
Employing the MVGC2 toolbox \cite{barnett2014mvgc}, we fitted a VAR(1) model to these time series using a Locally Weighted Regression (LWR) estimator, obtaining the matrix of the coefficients $A$ and the residual covariance $V$. 
Autocorrelation matrices $\Gamma_k$ can then be computed through a Lyapunov equation (Eq.~\eqref{eq:Lyapnuov}), and from here the full covariance of the system can be reconstructed. 
Interpreting the target of the decomposition $T$ as the joint future state of the VAR model, and the sources $X,Y$ as the past information of the system, we can calculate the total mutual information (TMI) %
as
\begin{equation} \label{eq:MI_var}
    I(X,Y;T) = \frac{1}{2} \log{|\Gamma_0|} - \frac{1}{2} \log{|V|} \, ,
\end{equation}
and proceed with the PID decomposition and any desired normalisation.
By repeating the process 1000 times for each subject and condition, we obtain a set of (possibly normalised) PID atoms that quantify the information flow between regions of the brain over time. 

\begin{figure}
    \centering
    \hspace*{-1.5mm}\includegraphics[width=0.5\textwidth]{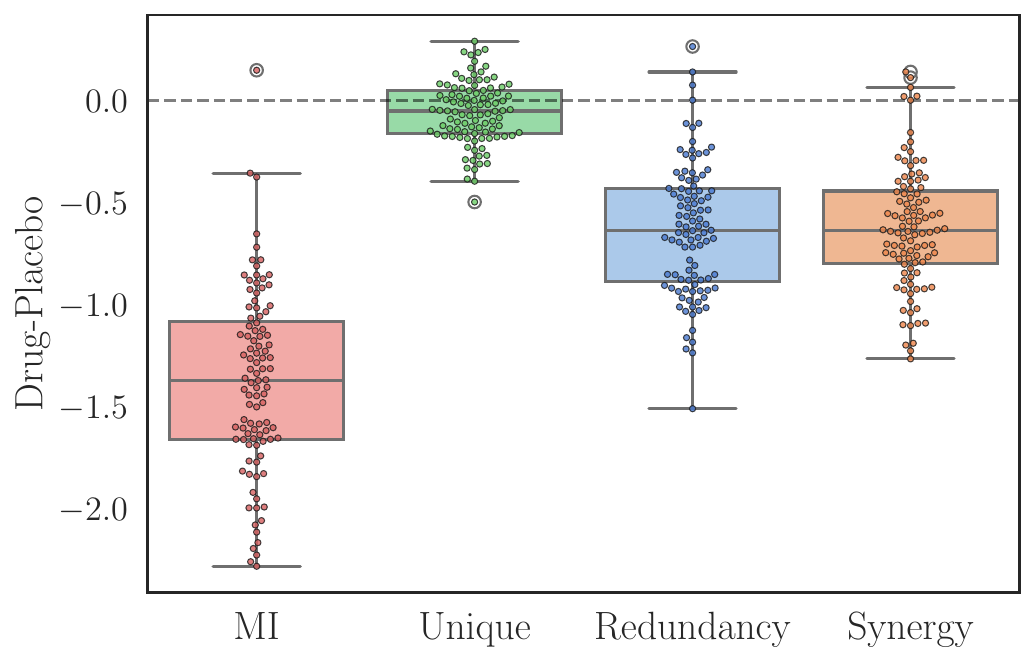}
    \caption{Raw values of PID atoms distribution for subject 1 under ketamine and placebo effects using the MMI PID.
    }
    \label{fig:S1_dp}
\end{figure}

As an example, in Fig.~\ref{fig:S1_dp} we present the resulting distribution of the PID components for the subject S1 under ketamine and placebo, focusing on the difference between atoms obtained in the drug and placebo conditions. 
It can be already noted that TMI is consistently higher in the placebo condition, thus potentially affecting the overall values of all PID atoms and indicating the necessity of a robust normalisation technique. 

Eventually, averaging over these points and repeating the same process for each subject gives the results presented in Figs.~\ref{fig:all_dp_MMI}, \ref{fig:all_dp_Idep}, and~\ref{fig:all_dp_Iccs}.

\subsection{NuMIT normalisation for VAR models} \label{sec:VAR_null} 

Here we develop the null model normalisation technique for the VAR model, which was employed for the neural analyses in Sec.~\ref{sec:neural}. 

From the defining evolution equation of the model (Eq.~\eqref{eq:VAR_eq3}), we note that the VAR model arises as a non-Markovian generalisation of the Gaussian system studied in Sec.~\ref{sec:pid_distr_study}, in the case in which the target variable is the future state of the sources ($d_S=d_T$). 
Therefore, in principle the parameters of the null systems could be sampled analogously to the Gaussian case (Sec.~\ref{sec:results_maths}).
However, due to the scaling properties of the Lyapunov equation (Eq.~\eqref{eq:Lyapnuov}), rescaling the noise covariance $V$ by a factor $g$ also affects the covariance matrices in the same way, thus leading to the same TMI. 
This property of the VAR model intrinsically satisfies one of our requirements for a normalisation procedure, as it avoids the problem of the PID atoms being dependent on the noise of the system. %
However, it is still necessary to embed a parameter in the model so that PID also becomes independent of the value of the system's mutual information.   
To achieve this, we repurpose $g$ as the spectral radius of the evolution matrices $A_k$, so that the autocovariances $\Gamma_k,\,k=0,\ldots,p$ become implicit functions of $g$.
In other words, as the maximum of the absolute values of the eigenvalues of $A_k$, the parameter $g$ determines the signal-to-noise ratio of the time series and therefore modulates the mutual information exchanged between sources and target.

Thus, once the TMI of the real system ${p(S,T)}$ is known, for each null model ${q_i(S,T)}$ we sample ${A_k^{ij} \sim \mathcal{N}(0,1)}$ and ${V \sim W_{d_S}(\mathbbb{1}_{d_S{\times}d_S},d_S)}$, then optimise $g$ through Eqs.~\eqref{eq:Lyapnuov}-\eqref{eq:MI_var} to satisfy the constraint on the TMI. 
Eventually, marginal mutual information can be calculated and PID decomposition performed.

\subsection{Multiplicative regression model} \label{sec:meth_mult_regr}

In Sec.~\ref{sec:res_mult_regr} we employed a regression model to assess the similarity between different PID definitions depending on the normalisation used. 
The goal was to observe how accurately a normalised PID atom computed with CCS can predict the corresponding normalised value computed with MMI, both with NMI and NuMIT. 
This was achieved by building a multiplicative regression model with two predictors, the CCS PID atoms $P_{\text{CCS}}$ --- obtained for each subject and normalised with either method --- and a binary dummy variable \textit{m}, set to 0 for the NMI normalisation and to 1 for NuMIT. Note that $P_{\text{CCS}}$ corresponds to the difference in the corresponding atom between drug and placebo conditions for a given subject, averaged across all sets of brain regions (similarly for MMI).
Mathematically,
\begin{equation} \label{eq:mult_regr}
    P_{\text{MMI}} = \beta_0 + \beta_1 \,P_{\text{CCS}} + \beta_2 \,\text{\textit{m}} + \beta_3\, P_{\text{CCS}} \,\text{\textit{m}} \, , 
\end{equation}
where $\beta_i, \, i=0,1,2,3$ are the regression coefficients. 
Considering the PID atoms for each subject obtained in Figs.~\ref{fig:all_dp_MMI} and \ref{fig:all_dp_Iccs}, we first standardise the points to mean 0 and variance 1, and then estimate the model parameters $\beta_i, \, i=0,1,2,3$ along with their p-values.
The $\beta_3$ term is of particular interest, as it quantifies the extent to which NuMIT normalisation ($m=1$) increases the correlation between $P_{\text{MMI}}$ and $P_{\text{CCS}}$. The fitted models are shown in Fig.~\ref{fig:mult_regr} and the p-values for $\beta_3$ in Table~\ref{tab:mult_regr_Iccs_MMI}.

We performed analogous analyses for the DEP-MMI and CCS-DEP comparisons reported in App.~\ref{app:mult_regr_app}.

\subsection{Neural Data} \label{sec:neural_data}
The data employed in the analysis consist of pharmaco-MEG recordings of patients under LSD \cite{carhart2016neural} (15 subjects),  psilocybin (PSI) \cite{muthukumaraswamy2013broadband} (14 subjects), and ketamine (KET) \cite{muthukumaraswamy2015evidence} (19 subjects) drugs in resting states. Data were obtained from an open data repository~\cite{megdata}.
We provide a short overview of the datasets here -- interested readers are referred to the original studies for an exhaustive description of the methods and the acquisition details of each experiment.

\subsubsection{Participants and drug infusion}

All participants gave informed consent to take part in the studies, approved by the UK National Health Service, and were excluded if they were younger than 21 years old, pregnant, had a personal or immediate family history of psychiatric disorder, suffered from substance dependence, or were smokers (only for KET). 
Moreover, subjects were exempted if they suffered from a medical condition that would render the volunteer unsuitable, such as psychiatric disorders, cardiovascular diseases, claustrophobia, blood or needle phobia, problematic alcohol abuse, and others. 
Also, patients undergoing LSD and PSIL delivery must have had previous experience with hallucinogenic drugs but not within 6 weeks of the study.

Drug delivery comprised of intravenous administration of a fixed single dose for LSD and PSIL, and a continued infusion of 40 minutes for KET.  
PSIL and KET data were obtained immediately after drug delivery, whereas for LSD the data were obtained after 4 hours, due to the slow pharmacodynamics of the drug. 
Placebo conditions involved an injection of saline solution and were conducted with identical procedures and under identical conditions to the corresponding drug.

\subsubsection{Data acquisition and preprocessing}
The data were recorded with a 271-gradiometer CTF MEG scan at the Cardiff University Brain Research Imaging Centre (CUBRIC). Each patient underwent two scanning sessions, both in eyes-closed resting state post-administration of drugs and placebo. 
Source-reconstruction of the data was performed on the centroid of the Automated Anatomical Labelling (AAL) brain atlas~\cite{tzourio2002automated} using a linearly constrained minimum variance beamformer~\cite{van1997localization}. Raw data was collected with a sampling frequency of 600Hz and split into 2-second epochs (i.e.\ 1200 timepoints). 
Preprocessing was performed with the FieldTrip toolbox \cite{oostenveld2011fieldtrip}, and consisted of the rejection of bad epochs, bad channels, and bad ICA components by visual inspection, a lowpass filter at 100Hz and a downsampling to 200Hz. Line noise was removed by fitting a sinusoidal signal at 50Hz and harmonics using a least-squares method, then subtracting it from the data.

\newpage

\twocolumngrid
\bibliographystyle{ieeetr}
\bibliography{bibliography}

\onecolumngrid
\clearpage

\appendix

\section{Remaining MEG results} \label{app:neural_app}
In this section we present the results of the analyses of psychedelic MEG data obtained with the $\text{DEP}$ and $\text{CCS}$ PID formulations.

\subsection{PID via dependency constraints ($\text{DEP}$)}

\begin{figure*}[ht]
    \centering
    \hspace*{-10mm}\includegraphics[width=1.1\linewidth]{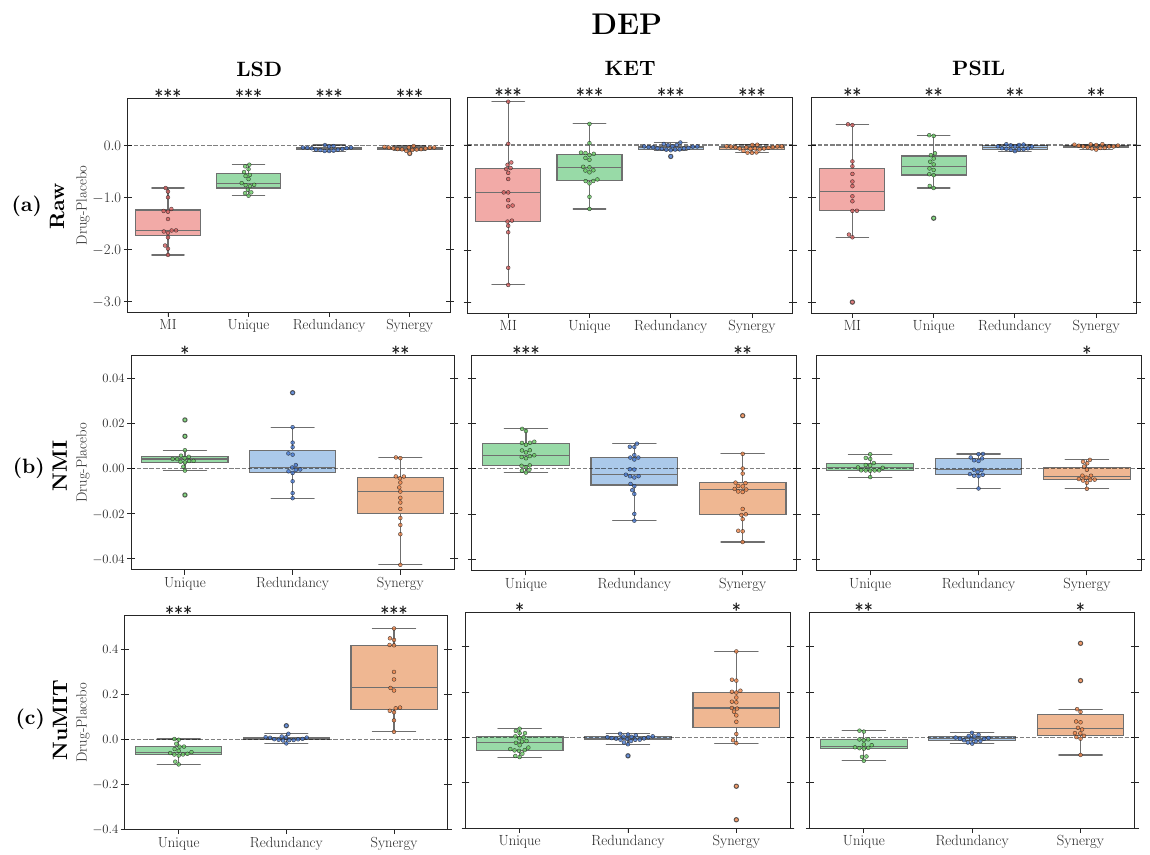}
\caption{PID atoms distribution for all subjects under different drugs and placebo effects using the DEP PID. From left to right, results refer to LSD, ketamine, and psilocybin drugs. Panel rows represent (a) the raw values of PID atoms, (b) the NMI-normalised PID atoms, and (c) the NuMIT-normalised PID atoms. The dashed black lines are drawn at zero. (P-values calculated with a one-sample t-test against the zero-mean null hypothesis. *: $p<0.05$; **: $p<0.01$; ***: $p<0.001$).}
\label{fig:all_dp_Idep}
\begin{subfigure}{0.45\textwidth}
    \phantomcaption
    \label{fig:idep_dp_abs}
\end{subfigure}
\hfill
\begin{subfigure}{0.45\textwidth}
    \phantomcaption
    \label{fig:idep_dp_nmi}
\end{subfigure}
\hfill
\begin{subfigure}{0.45\textwidth}
    \phantomcaption
    \label{fig:idep_dp_null}
\end{subfigure}
\end{figure*}

Unique Information via Dependency Constrains ($\text{DEP}$) \cite{james2018unique} is a PID formulation in which the unique information is quantified by a statistical dependency decomposition, i.e.\ a procedure that assesses how the dependencies between the sources affect the mutual information between sources and targets. 
In Fig.~\ref{fig:all_dp_Idep} we present the results of the MEG analysis obtained by using the DEP definition of unique information.

As expected, the higher TDMI found in the placebo condition leads to a decrease in the raw values of all PID atoms under the effects of every drug. 
While NMI yields results inconsistent across drugs, the results obtained by NuMIT normalisation show consistently higher synergistic values across all three psychedelics.

Finally, note that although raw (Fig.~\ref{fig:idep_dp_abs}) and NMI-normalised (Fig.~\ref{fig:idep_dp_nmi}) PID values show contrasting behaviour with the MMI results presented in the main text (Sec.~\ref{sec:neural}), the outcomes of the null model procedure are broadly consistent between different PID measures.

\subsection{PID via common change in surprisal ($\text{CCS}$)}

In Redundant Information with Pointwise Change in Surprisal ($\text{CCS}$) PID, redundancy is quantified by specific local co-information terms drawn from a maximum entropy distribution, providing an intuitive interpretation while also satisfying the core axioms of a redundancy metric, finding broad applications in various areas \cite{ince2017measuring}. 

\begin{figure*}[ht]
    \centering
    \hspace*{-10mm}\includegraphics[width=1.1\linewidth]{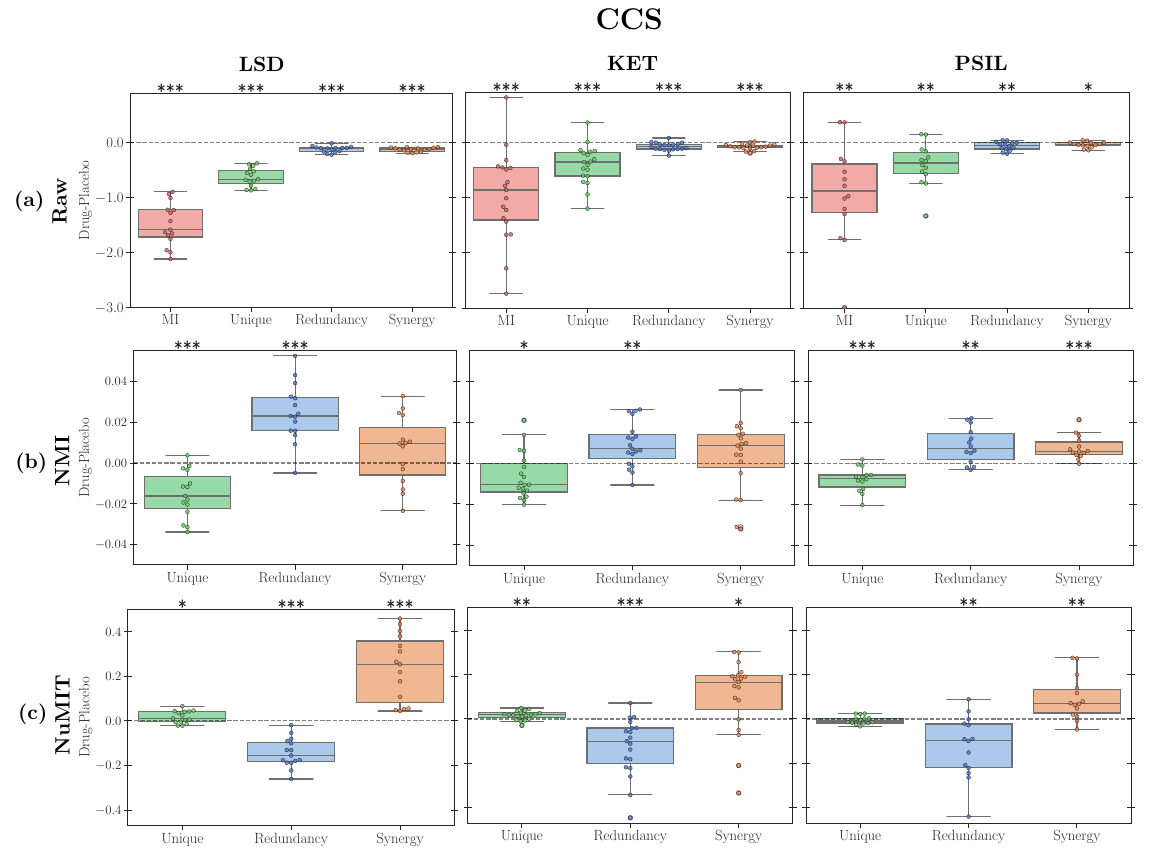}
\caption{PID atoms distribution for all subjects under different drugs and placebo effects using the CCS PID. From left to right, results refer to LSD, ketamine, and psilocybin drugs. Panel rows represent (a) the raw values of PID atoms, (b) the NMI-normalised PID atoms, and (c) the NuMIT-normalised PID atoms. The dashed black lines are drawn at zero. (P-values calculated with a one-sample t-test against the zero-mean null hypothesis. *: $p<0.05$; **: $p<0.01$; ***: $p<0.001$).}
\label{fig:all_dp_Iccs}
\begin{subfigure}{0.45\textwidth}
    \phantomcaption
    \label{fig:iccs_dp_abs}
\end{subfigure}
\hfill
\begin{subfigure}{0.45\textwidth}
    \phantomcaption
    \label{fig:iccs_dp_nmi}
\end{subfigure}
\hfill
\begin{subfigure}{0.45\textwidth}
    \phantomcaption
    \label{fig:iccs_dp_null}
\end{subfigure}
\end{figure*}

Using this formulation, we obtain the results shown in Fig.~\ref{fig:all_dp_Iccs}.

As before, the PID atom distributions calculated with the null normalisation technique appear to be strongly indicative of higher synergistic contributions in the information flow of brains under all three psychoactive substances.  

\subsection{Multiplicative regression for $\text{DEP}-$MMI and $\text{CCS}-\text{DEP}$} \label{app:mult_regr_app}

\begin{figure*}[h]
    \centering
    \hspace*{-10mm}\includegraphics[width=1.1\linewidth]{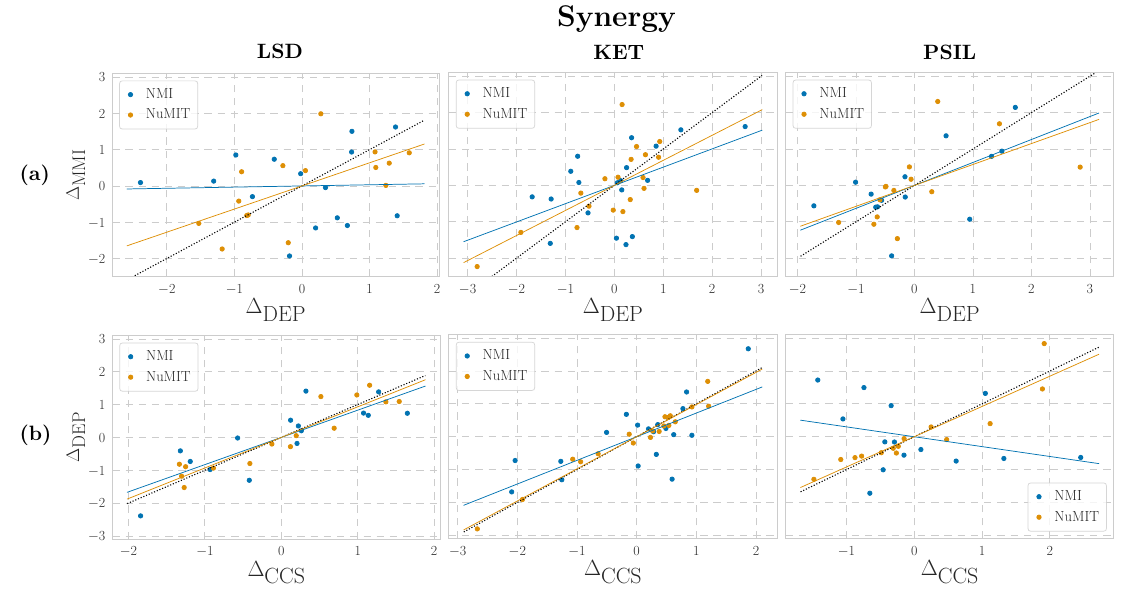}
\caption{Multiplicative regression models for NMI- and NuMIT-normalised synergies between (a) $\text{DEP}$ and MMI (b) $\text{CCS}$ and $\text{DEP}$ PID definitions, for LSD, ketamine, and psilocybin.  $\Delta$ denotes the differences between drug and placebo in PID atoms obtained with any PID (MMI, CCS, or DEP). Black dotted line is the bisector of the first quadrant.}
\label{fig:mult_regr_idep_iccs_mmi}
\begin{subfigure}{0.45\textwidth}
    \phantomcaption
    \label{fig:mult_regr_lsd_iccs_mmi}
\end{subfigure}
\hfill
\begin{subfigure}{0.45\textwidth}
    \phantomcaption
    \label{fig:mult_regr_lsd_iccs_mmi_red}
\end{subfigure}
\end{figure*}

Here we present the comparisons between DEP vs.\ MMI and CCS vs.\ DEP PID definitions with the multiplicative regression model introduced in Sec.~\ref{sec:neural}, with appropriate changes to the atoms in Eq.~\eqref{eq:mult_regr}. 
Since redundancies and unique information follow different definitions and interpretations in these formulations, we only focus on the comparisons between synergy atoms, performing the regression and presenting the results in Fig.~\ref{fig:mult_regr_idep_iccs_mmi}.

Overall, we see that in both $\text{DEP}$-MMI and $\text{CCS}$-$\text{DEP}$ cases NuMIT performs better than NMI, except for psilocybin in the DEP-MMI case (top right of Fig.~\ref{fig:mult_regr_lsd_iccs_mmi}), in which both normalisations behave very similarly. 
Analogously to the results in the main text, we present in Tabs.~\ref{tab:mult_regr_Idep_MMI} and \ref{tab:mult_regr_Iccs_Idep} the slopes of the regression lines, i.e.\ the Pearson coefficient of the PID atoms, as well as the p-value of the interaction coefficients $\beta_3$ of Eq.~\eqref{eq:mult_regr}. 

These results, along with the ones obtained in Sec.~\ref{sec:neural}, support the idea that the null model normalisation might be able to capture the emergent components of the system more deeply, as it yields synergy atoms that more consistent across different PID definitions.

\renewcommand{\arraystretch}{1.2}
\begin{table}[h]
\begin{tabular}%
    {
    >{\centering}p{0.2\columnwidth} 
    >{\centering}p{0.2\columnwidth}
    >{\centering}p{0.15\columnwidth}
    >{\centering}p{0.15\columnwidth}
    >{\centering}p{0\columnwidth} 
    >{\centering\arraybackslash}p{0.15\columnwidth}
    }
\multirow{2}{*}{PID atom} & \multirow{2}{*}{Drug} & \multicolumn{2}{c}{Correlations} &  & \multirow{2}{*}{p-value} \\ \cline{3-4}
&  & NMI & NuMIT &  &  \\ \hline
\multirow{3}{*}{Synergy} 
& LSD & 0.032 & 0.638 &  & 0.10 \\
& KET & 0.502 & 0.689 &  & 0.50 \\
& PSIL & 0.631 & 0.575 &  & 0.87 \\ \hline
\end{tabular}
\caption{Pearson correlations and $\beta_3$ p-values of the multiplicative regression model of Eq.~\eqref{eq:mult_regr} (replacing $P_{\text{CCS}}$ with $P_{\text{DEP}}$) for synergy ($\text{DEP}$-MMI comparison).}
\label{tab:mult_regr_Idep_MMI}
\end{table}

\renewcommand{\arraystretch}{1.2}
\begin{table}[]
\begin{tabular}%
    {
    >{\centering}p{0.2\columnwidth} 
    >{\centering}p{0.2\columnwidth}
    >{\centering}p{0.15\columnwidth}
    >{\centering}p{0.15\columnwidth}
    >{\centering}p{0\columnwidth} 
    >{\centering\arraybackslash}p{0.15\columnwidth}
    }
\multirow{2}{*}{PID atom} & \multirow{2}{*}{Drug} & \multicolumn{2}{c}{Correlations} &  & \multirow{2}{*}{p-value} \\ \cline{3-4}
&  & NMI & NuMIT &  &  \\ \hline
\multirow{3}{*}{Synergy} 
& LSD & 0.830 & 0.931 &  & 0.59 \\
& KET & 0.720 & 0.978 &  & 0.15 \\
& PSIL & -0.300 & 0.921 &  & 0.004 \\ \hline
\end{tabular}
\caption{Pearson correlations and $\beta_3$ p-values of the multiplicative regression model of Eq.~\eqref{eq:mult_regr} (replacing $P_{\text{MMI}}$ with $P_{\text{DEP}}$) for synergy ($\text{CCS}$-$\text{DEP}$ comparison).}
\label{tab:mult_regr_Iccs_Idep}
\end{table}

\FloatBarrier

\section{Further Noise Sweep analysis} \label{app:noise_sweep}

In this section, we present further analyses of our proposed normalisation method on the Gaussian system of Eq.~\eqref{eq:gauss_pid_distr}. 
In particular, verifying the independence of the normalised atoms of the noise of the system for an additional case of MMI, as well as using DEP and CCS definitions.

\begin{figure}[h]
    \centering
    \includegraphics[width=0.5\columnwidth]{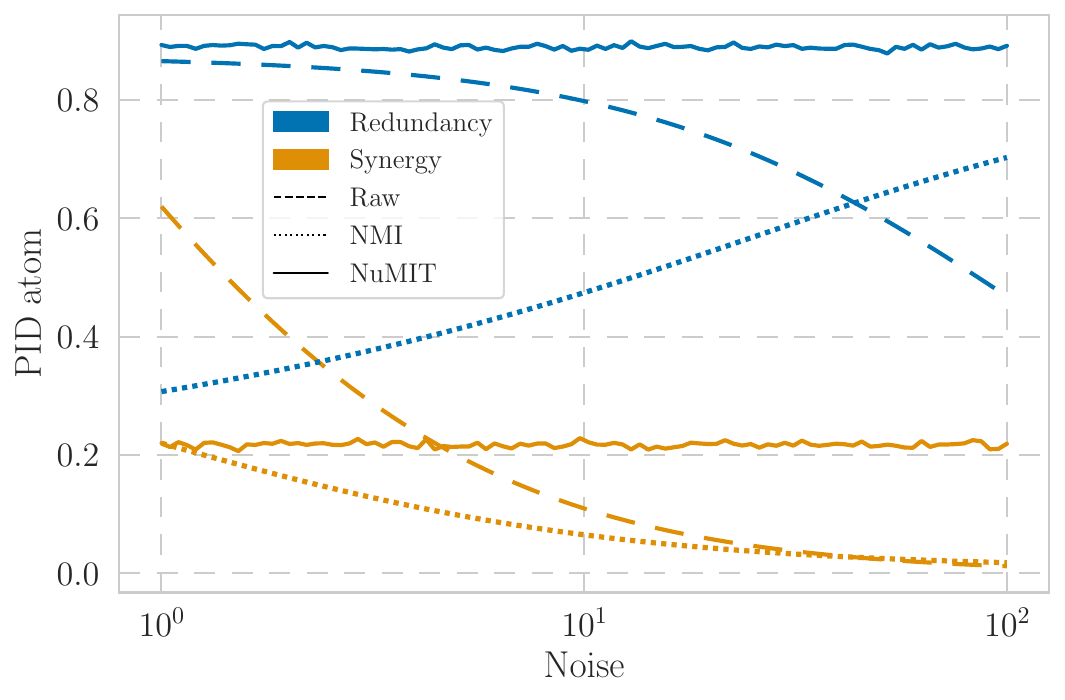} 
    \caption{Raw, NMI and Null-model redundancy and synergy atoms calculated for an asymmetric system given by Eqs. \eqref{eq:gauss_pid_distr} and \eqref{eq:Gaussian_parameters_asym} for noise level $g\in[1,100]$ and MMI definition.}
    \label{fig:NMI_fail_null_asym}
\end{figure}

We begin by studying the MMI PID. As the system considered in Eq.~\eqref{eq:Gaussian_parameters} of Sec.~\ref{sec:keyint_norm} is highly symmetric and might be hiding non-trivial dependencies between the sources, we perform the same noise sweep procedure shown in Fig.~\ref{fig:keyint_NMI} with the following randomly sampled parameters
\begin{equation} \label{eq:Gaussian_parameters_asym}
    A = \begin{pmatrix} -0.3 & 0.9 \end{pmatrix} 
        \quad
    \Sigma_S = 50 \begin{pmatrix}
                5 & 7\\
                7 & 11
                \end{pmatrix} \,,
\end{equation}
Examining the results shown in Fig.~\ref{fig:NMI_fail_null_asym}, we observe again that the null-normalised atoms assume a constant value for different values of noise, whereas the NMI ones vary. 

\begin{figure*}[h]
    \centering
    \hspace*{-10mm}\includegraphics[width=1.1\linewidth]{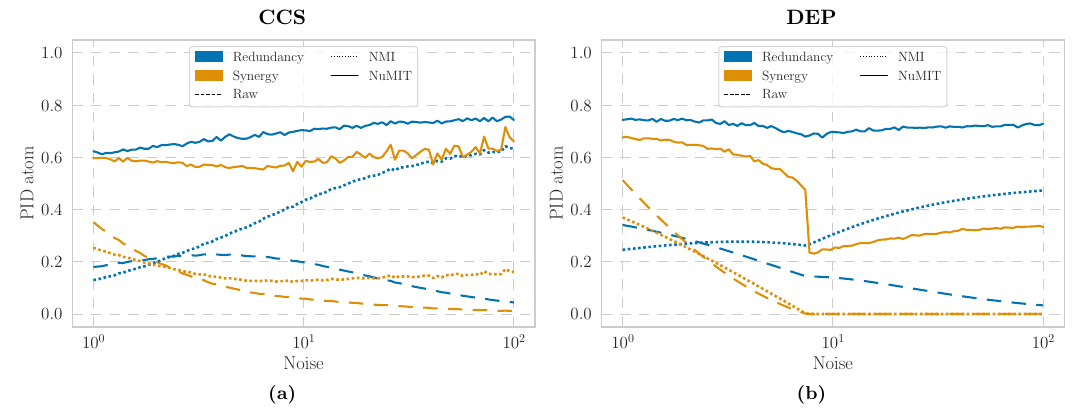}
\caption{Raw, NMI and Null-model redundancy and synergy atoms calculated for an asymmetric system given by Eqs. \eqref{eq:gauss_pid_distr} and \eqref{eq:Gaussian_parameters} for noise level $g\in[1,100]$ and (a) $\text{CCS}$ and (b) $\text{DEP}$ definitions.}
\label{fig:nmi_fail_idep_iccs}
\begin{subfigure}{0.45\textwidth}
    \phantomcaption
    \label{fig:nmi_fail_iccs}
\end{subfigure}
\hfill
\begin{subfigure}{0.45\textwidth}
    \phantomcaption
    \label{fig:nmi_fail_idep}
\end{subfigure}
\end{figure*}

Turning to other PID definitions, we now test our model performing the same procedure with $\text{DEP}$ and $\text{CCS}$ formulations. Using the parameters of Eq.~\eqref{eq:Gaussian_parameters}, we obtain the results reported in Fig.~\ref{fig:nmi_fail_idep_iccs}. We observe that for $\text{CCS}$, synergy remains nearly constant, while redundancy exhibits gradual growth. Nonetheless, its increase is significantly smaller compared to that of the NMI redundancy.

On the other hand, in the $\text{DEP}$ case, although redundancy shows near-flat behaviour, the null-normalised synergy presents a sudden drop when its raw value becomes zero. 
Although this is not the desired behaviour, it is not surprising that it happens when the PID component vanishes: as our model is based on the assessment of percentiles, the calculations are sensitive to small variations in the distributions, especially when dealing with small numbers, and potentially result in unstable estimates. 
Moreover, we underline that DEP formulation entails a minimisation procedure over the edges of a constraint lattice, which can indeed lead to discontinuous jumps when the TMI of the system changes.

Thus, caution is recommended when a PID atom value becomes zero. From a practical point of view, it might be more meaningful to study these cases separately without the need for a normalisation technique.

\FloatBarrier

\section{Higher-dimensional systems} \label{app:higher_dims}

Except for the VAR model results on neural data, so far the operation of the null model has been shown for univariate systems. In this section we briefly show that our procedure can be easily generalised to multivariate sources, producing the desired results.

As information-theoretic quantities depend on the dimensionality of the random variables involved, we expect the shapes of the PID-atom distributions shown in Fig.~\ref{fig:S2T1_all_PID} to change with the dimensionality of the sources (note that in all cases we perform a two-source PID, with multivariate sources). 
Considering the Gaussian system of Eq.~\eqref{eq:gauss_pid_distr_mult}, we repeat the same procedure followed in Sec.~\ref{sec:pid_distr_study} and analyse the average value of the null distributions for various dimensions $d_S$ of the multivariate sources. 
\begin{figure*}[ht]
    \centering
    \hspace*{-10mm}\includegraphics[width=1.1\linewidth]{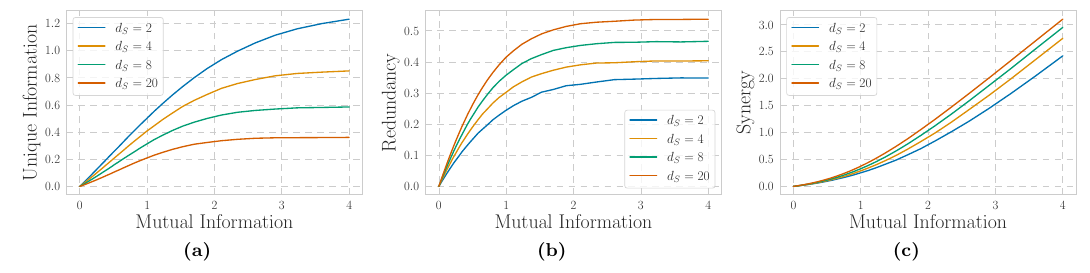}
\caption{Averages of PID atoms for different values of mutual information compared across Gaussian systems of various sizes for (a) redundancy, (b) unique information, (c) synergy.
}
\label{fig:MI_sweep_higher_dims}
\begin{subfigure}{0.45\textwidth}
    \phantomcaption
    \label{fig:all_av_sizes}
\end{subfigure}
\hfill
\begin{subfigure}{0.45\textwidth}
    \phantomcaption
    \label{fig:All_averages_redundancy}
\end{subfigure}
\hfill
\begin{subfigure}{0.45\textwidth}
    \phantomcaption
    \label{fig:All_averages_synergy}
\end{subfigure}
\end{figure*}%
The results reported in Fig.~\ref{fig:MI_sweep_higher_dims} depict a clear trend:  synergy and redundancy increase for larger systems, while unique information becomes smaller. %
Moreover, one can notice that the non-linearity of their behaviour is accentuated by a larger system size.

\begin{figure*}[ht]
    \centering
    \hspace*{-10mm}\includegraphics[width=1.1\linewidth]{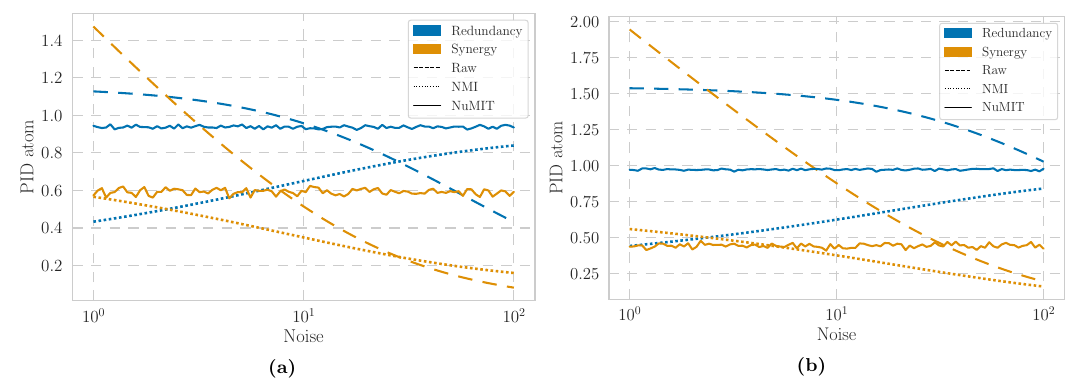}
\caption{Raw, NMI and Null-model redundancy and synergy atoms calculated for the system given by Eq. \eqref{eq:gauss_pid_distr_mult} for noise level $g\in[1,100]$ and (a) $d_S=8$ and (b) $d_S=20$. Parameters $A, \Sigma_S, \Sigma_{\epsilon}$ were randomly sampled.}
\label{fig:nmi_fail_SS}
\begin{subfigure}{0.45\textwidth}
    \phantomcaption
    \label{fig:nmi_fail_s8}
\end{subfigure}
\hfill
\begin{subfigure}{0.45\textwidth}
    \phantomcaption
    \label{fig:nmi_fail_s20}
\end{subfigure}
\end{figure*}

Turning now to the null model implementation, we test the robustness of our method in the multivariate case by performing the usual noise sweep procedure on systems of size $d_S=8$ and $d_S=20$ (Fig.~\ref{fig:nmi_fail_SS}). 
Once again, the atoms normalised with the null model are approximately independent of the noise strength, while raw and NMI-normalised values vary greatly.

\FloatBarrier

\section{Null-model normalisation for discrete random variables} \label{app:discrete}

Finally, we suggest a possible implementation of our technique for discrete random variables. We first present the mathematical formulation (Sec.~\ref{sec:discrete_null}), then validate the model on synthetic systems in Sec.~\ref{sec:discrete_validation}.

\subsection{Normalisation for discrete variables} \label{sec:discrete_null}

Let $X$ and $Y$ be two discrete variables jointly sampled from a multinomial distribution Mult$(n,k,p_1,...p_k)$, where $n$ is the number of trials, $k$ the number of possible events, and $p_1,...p_k$ the probabilities associated to each event, and take $f$ a Boolean function of $X$ and $Y$, e.g.\ a two-input logic gate. We define the target $T$ as
\begin{gather} \label{eq:discrete_eq}
    T = Z + \epsilon \, ,
\end{gather}
with $ Z = f(X;Y)$, and where $\epsilon$ is a discrete noise term that flips the value of $Z$ with probability $p_{\epsilon}$. 
The probabilities $p_1,...,p_k$ that govern the statistics of $X,Y$ can be sampled from a Dirichlet distribution Dir$(\alpha_1, ..., \alpha_k)$ defined on a $k-1$ simplex. For simplicity, we take $\alpha_i = \alpha$ $\forall i=1,...,k$. 
The entropies can be computed using the definition in Eq.~\eqref{eq:def_entropy}:
\begin{align} \label{eq:discr_entropies}
    H(T) = & \begin{aligned}[t]
    & -p(T=0)\log{p(T=0)}+ \\
    & -p(T=1)\log{p(T=1)} \, ,
\end{aligned} \\ \label{eq:discr_entropies2}
    H(T|S) = & \begin{aligned}[t]
    & -p_{\epsilon}\log{p_{\epsilon}}-(1-p_{\epsilon})\log{(1-p_{\epsilon})} \, ,\\
\end{aligned}
\end{align}
where $p(T=0) = p(Z=0)(1-p_{\epsilon})+p(Z=1)p_{\epsilon}$, and analogous for $p(T=1)$.
Once the function $f$ is chosen, mutual information can be calculated as the difference of Eqs.\ \eqref{eq:discr_entropies} and \eqref{eq:discr_entropies2} (see Eq.~\eqref{eq:MI_entr_def} in Sec.~\ref{sec:methods}). 

The construction of the null model is then analogous to the cases already described: sampling probabilities ${p_1,...,p_k}$ from Dir($\alpha$) and choosing a random gate $f$, we treat $p_{\epsilon}$ as the parameter used for the optimisation, tuning it to produce the desired value of MI. 
Finally, marginal mutual information $I(X;T)$, $I(Y;T)$ can be computed from the definition of entropy, and PID is performed. 
In the following section, we provide a practical example of the procedure.

\subsection{Validation in synthetic systems} \label{sec:discrete_validation}

We illustrate here the application of the null model normalisation on discrete systems of the form given by Eq.~\eqref{eq:discrete_eq}, in which $f$ is a 2-input logic gate.

Logic gates are a class of discrete systems that perform Boolean operations on binary inputs, being often used to study the behaviour of information-theoretic quantities in simple cases \cite{ince2017measuring}. 
In our perspective, we consider binary inputs $X$ and $Y$, whose probability is sampled from a Dirichlet distribution, employing 2-input logic gates as proxies to test the functioning of the null model. 
Analogously to the Gaussian system in Sec.~\ref{sec:synthetic}, we focus on maximal synergistic, redundant, and unique information systems, performing the null model procedure and expecting an atom quantile close to the unit value. 

Since the number of logic gates for binary variables is relatively small ($2^4=16$), we construct the null model considering all of such systems. 
However, of those 16, we can neglect both the tautology and contradiction, as the output is independent of the input and therefore $I(T;X,Y)=0$, and those gates which are the same under the symmetry $0\leftrightarrow1$, as it leaves the statistical structure of the model invariant. 
Hence, we are left with only 7 gates which are reported in Tab.~\ref{tab:gates}; among those, we can recognise the XOR ($Z_1$) and the OR ($Z_4$) gates. 

\setlength{\tabcolsep}{10pt}
\begin{table}
\begin{tabular}{cc|c|c|c|c|c|c|c}
    $X$ & $Y$ & $Z_1$ & $Z_2$ & $Z_3$ & $Z_4$ & $Z_5$ & $Z_6$ & $Z_7$ \\ \hline
    0 & 0 & 0 & 0 & 0 & 0 & 1 & 1 & 1 \\
    0 & 1 & 1 & 0 & 1 & 1 & 0 & 1 & 1 \\
    1 & 0 & 1 & 1 & 0 & 1 & 1 & 0 & 1 \\
    1 & 1 & 0 & 1 & 1 & 1 & 1 & 1 & 0 \\ \hline
\end{tabular}
\caption{2-input logic gates used for the null model.}
\label{tab:gates}
\end{table}

In Sec.~\ref{sec:discrete_null} we showed how the mutual information between source and targets can be computed. 
For the marginal mutual information $I(X;T)$ and $I(Y;T)$, we make use of 
\begin{equation}
    I(X;T) = H(T) - H(T|X) \,,
\end{equation}
recalling the definition of entropy and conditional entropy Eqs.~\eqref{eq:def_entropy} and \eqref{eq:cond_entr_def}, and manually calculating their values using the gate definitions of Tab.~\ref{tab:gates}. 
After an analogous procedure is performed for $I(Y;T)$, the PID decomposition is applied. \\

\begin{figure*}[ht]
    \centering
    \hspace*{-10mm}\includegraphics[width=1.1\linewidth]{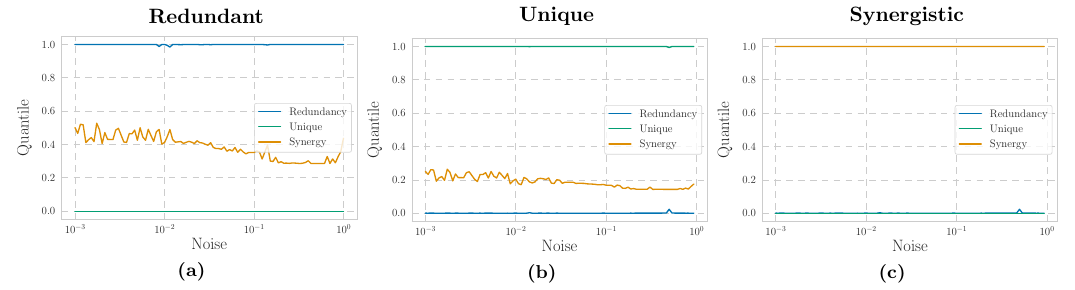}
\caption{Quantiles of the PID atoms for discrete models of Eq.~\eqref{eq:discrete_eq}  for various noise levels $p_{\epsilon}\in[0.001,1)$. (a) Predominantly redundant system ($Z_5$), (b) predominantly unique system ($Z_2$), (c) predominantly synergistic system ($Z_1$).}
\label{fig:max_pid_sweep_discr}
\begin{subfigure}{0.45\textwidth}
    \phantomcaption
    \label{fig:max_red_discr}
\end{subfigure}
\hfill
\begin{subfigure}{0.45\textwidth}
    \phantomcaption
    \label{fig:max_un_discr}
\end{subfigure}
\hfill
\begin{subfigure}{0.45\textwidth}
    \phantomcaption
    \label{fig:max_syn_discr}
\end{subfigure}
\end{figure*}

Analogous to Sec.~\ref{sec:synthetic}, we analyse whether the null normalisation is robust to noise changes. 
Starting with a predominantly redundant system, we take the table of truth given by $Z_5$, with probabilities $p_{00}=p_{11}=\delta/2$, $p_{01}=p_{10}=(1-\delta)/2$, where $\delta$ is a small positive number (0.01) and $p_{ab}$ indicates the joint probability $p(X=a,Y=b)$. 
This way we obtain a system in which $X$ and $Y$ provide the same amount of information due to symmetry, with minimal synergistic component as $\{00\}$ and $\{11\}$ inputs are suppressed by the low probability. 
For each level of noise, we perform the decomposition and run the null model over the space of the 7 possible logic gates, obtaining a distribution of redundancy atoms and the quantiles shown in Fig.~\ref{fig:max_red_discr}. 
As expected, we observe a high quantile ($\sim 1$) for the maximally redundant system, independently of the noise level. Although synergy atom shows some fluctuations, redundancy is clearly greater than the other two atoms throughout. 

Correspondingly, as a predominantly unique system we consider the $Z_2$ gate with equal probabilities $p_{00}=p_{01}=p_{10}=p_{11}=0.25$. 
In fact, if not for the possible flip induced by the noise term $\epsilon$, the knowledge of $X$ would completely determine the output T. 
Spanning $p_{\epsilon}\in[0.001,1)$ and proceeding with the null model construction, we report the atoms' behaviour in Fig.~\ref{fig:max_un_discr}, verifying the expected trend.

At last, we turn to a purely synergistic system, i.e.\ a model in which the knowledge of one variable does not convey any information on the outcome of the target. This is obtained considering the XOR ($Z_1$) with independent probabilities $p_{00}=p_{01}=p_{10}=p_{11}=0.25$.
Following the same steps as above we obtain Fig.~\ref{fig:max_syn_discr}, confirming the expected results.

\end{document}